\documentclass[aps,superscriptaddress,showpacs,nofootinbib]{revtex4}
\usepackage{amsmath}
\usepackage{graphicx}
\usepackage{epsfig}
\newcommand{\be}{\begin{equation}}
\newcommand{\ee}{\end{equation}}
\newcommand{\bea}{\begin{eqnarray}}
\newcommand{\eea}{\end{eqnarray}}
\newcommand{\ud}{\mathrm{d}}

\begin{document}

\begin{flushright}
UGFT-268/10 \\
CAFPE-138/10 \\
Edinburgh 2010/10
\end{flushright}

\title{Power suppression  from disparate mass scales in effective scalar field 
theories of inflation and quintessence}

\author{Mar~Bastero-Gil}
\email{mbg@ugr.es}
 
\affiliation{
Departamento de F\'{\i}sica Te\'orica y del Cosmos,
Universidad de Granada, Granada-18071, Spain}%

\author{Arjun~Berera}
\email{ab@ph.ed.ac.uk}
\affiliation{
School of Physics and Astronomy, University of Edinburgh, Edinburgh EH9 3JZ, UK}

\author{Brendan M. Jackson}
\email{bmj@roe.ac.uk}
\affiliation{Institute for Astronomy, School of Physics and Astronomy,
University of Edinburgh, Edinburgh EH9 3HJ, UK}


\date{\today}

\begin{abstract}
A scalar potential coupled to other fields of large disparate
masses will exhibit power suppression of the quantum loop corrections
from these massive fields.  Quintessence fields in the dark energy regime 
and inflaton fields during inflation often 
have a very large background field value. Thus any other field
with its mass dependent on 
the quintessence/inflaton background field value through a moderate
coupling will become
very massive during the dark energy/inflation phase and its quantum corrections
to the scalar effective potential will be suppressed.
This concept is
developed in this paper using the decoupling theorem.  The problem
then reduces to a quantitative question of the size of suppression
effects within the parameter space of coupling constants, scalar field 
background value and renormalization scale.  Some numerical
examples are presented both
for inflation and quintessence, but the approach is general 
and can be applied to any scalar field effective potential.  
The consequences to dark energy of 
the decoupling  effect developed here is that the quintessence field need 
not just be an incredibly weakly interacting field, often added as simply 
an add-on to generate dark energy and have no other purpose.
Instead, this quintessence field could play a central role in the 
particle physics dynamics at early times and then simply decouple at late 
times before the onset of the dark energy phase.
For inflation a consequence is coupling of the inflaton to other
fields can be much larger in certain models, without needing
supersymmetry to control quantum corrections.
\end{abstract}


\pacs{11.10.Hi, 11.10.Gh, 98.80.Cq}
\keywords{}


\maketitle

\section{Introduction}

Often in cosmology, scalar fields with appropriate
potentials and interactions are studied in order to explain different physical
effects. Very well known examples are inflation models, which provide
a solution for the horizon and flatness problems of the standard
cosmology, and generate a nearly scale invariant primordial density
perturbation, which has been tested by the observations of the cosmic
microwave background radiation spectrum \cite{wmap}. The simpler examples are
large field models of inflation with a renormalizable potential, a
mass term plus a quartic interaction \cite{ci}.  When the quartic
self-interaction dominates, the WMAP normalization of the primordial
spectrum demands this coupling to be tiny, $\lambda \simeq 10^{-14}$.
Present observational data also indicate that we live in an accelerated
expanding Universe today \cite{sn1a}, and that around 70\% of the
total energy  density is made of a component, called dark energy,  with negative
equation of state $w$, close to -1 \cite{sdss,wmap}. This could be
explained by a light rolling scalar field, usually called
quintessence \cite{quintessencep,Copeland:2006wr}. The dynamics of the
field is such that at early times, during matter or radiation
domination, its energy 
density $\rho_{DE}$ is subdominant, and it is only today, when the field
finds itself evolving in a roughly constant potential, that
$\rho_{DE}$ dominates. Commonly,
quintessence potentials are given by 
either  exponentials or inverse powers of the field,
i.e, by a non-normalizable potential,  
which yields naturally a small
classical mass of the quintessence field, below the Hubble parameter,
without the need of fine-tuning.  

In trying to explain the origin of the inflation and 
quintessence potential from a
particle physics model, it is unlikely that it will appear as an
isolated entity with no interactions to any other degree 
of freedom. For example, inflation should be follow by a reheating
period, such that we recover a radiation dominated universe at the end
of it. The standard picture is that the inflaton couples to
other light degrees of freedom into which it decays during
reheating \cite{reheatu}. 
Moreover in warm inflation dynamics \cite{Berera:1995ie}
interaction of the inflaton field with other fields is needed
to produce radiation concurrently with inflationary expansion.
For models where the inflaton has sizable coupling to other fields,
in order to maintain the required flatness of the potential,
supersymmetry is typically used to control large
quantum corrections \cite{Lyth:1998xn}.
For quintessence models, 
there are viable models
coupled to the dark matter fluid \cite{couplequint}. 
Or quintessence fields can directly be coupled  
through standard renormalizable interactions to
other bosons and fermions, providing for example time-varying masses
for these degrees of freedom \cite{couplequint2}.  But once coupled to
other species, one should check the stability of the classical results 
against quantum corrections. 
  
These sources of quantum corrections have been already studied in the
literature. For inflation for fields interacting sizably with
the inflaton, their corrections usually must be
controlled by supersymmetry. Often these effects are not a
problem, but indeed they can either help with the
inflationary trajectory, as in supersymmetric hybrid model
\cite{Dvali:1994ms}, or lead to different observational signatures
\cite{NeferSenoguz:2008nn}.  For quintessence, in
Ref. \cite{Brax:1999yv} the one-loop corrections due to the
self-interactions, among other cases, were studied by regularizing the
theory with a cut-off $\Lambda$, and they conclude that for reasonable
values of the high-energy cut-off, $\Lambda < m_P$, quantum corrections
do not spoil the quintessence potential. At early times, when the
quintessence energy density is still subdominant, quantum corrections
can be quite large compared to the tree-level potential, but still
subdominant compared to the other components of the energy density. By
the time the quintessence epoch starts, quantum corrections have
already become negligible. A similar conclusion is reached in
Ref. \cite{Garny:2006wc} for the self-couplings. More severe are the
constraints on renormalizable couplings to fermions and scalars, which
have to be really tiny and negligible to avoid distortion of the
quintessence potential \cite{Garny:2006wc,Doran:2002qd,Arbey:2007vu}.

At first glance it seems that the previous studies on quantum
corrections to the quintessence potential practically forbid couplings
to other scalars and/or fermions.  For inflation it appears that in order
to have sizable interaction with other fields requires supersymmetry.
Here we want to argue against these
conclusions, and show that quantum corrections 
(at one and two loop orders) can
be kept under control once the decoupling theorem is taken into
account \cite{decoupling}. 
The result does not depend on whether we study a quintessence
or inflation model.  As such we will deal with a generic scalar potential
$V(\phi)$, where the field has a non vanishing background field
value. We will focus on the interactions to scalars, by
introducing another scalar $\chi$ with potential: 
\be \label{v2f} 
V(\chi) =
\frac{1}{2} m_{\chi}^2 \chi^2 + \frac{\lambda_{\chi}}{4!} \chi^4 +
\frac{1}{2} g^2 \chi^2 \phi^2 \,.  
\ee 
The potentially dangerous term
is $g^2 \phi^2 \chi^2$ , which can give rise to a large 
quantum contribution
to the effective potential. This coupling induces
a large field dependent mass for the $\chi$ field, $m_\chi \sim g
\phi$, and when $\phi \sim m_P$, which is often the magnitude
during inflation \cite{ci,Lyth:1998xn,Linde:1993cn,ni} 
or quintessence \cite{quintessencep},
this can be larger than $\rho^{1/4}$,
$\rho$ being the total energy density. However, if we do not have
enough energy to excite such heavy states, physically we can expect
them to decouple from the spectrum \cite{decoupling}, and their
contribution to the effective potential to be highly suppressed.  To
properly address this issue, one must compute the effective potential
using renormalised perturbation theory in which decoupling is already
implemented. This is a well studied problem
\cite{bando2,nakano,casas}, and we want to examine the implications of
their results in the context of scalar field inflation and
quintessence.

The immediate problem in computing radiative corrections to the
effective potential is that the latter must be renormalization scale
independent; the point is how to choose that scale.  One
common approach is such that
all log corrections due to different mass scales are kept small
and can be resumed.  For models with a single mass scale
scale, the perturbative quantum corrections lead to
logarithmic terms of the form $\ln^n(m^2/\mu^2)$ and the standard
procedure for controlling the large log-terms is to choose the
renormalization scale near the mass scale $\mu \sim m$.
For a multi-mass case, such as in Eq. (\ref{v2f}), logarithmic
corrections terms will arise of the form
$\sim \ln^n(m_{\phi}^2/\mu^2) \ln^m(m_{\chi}^2/\mu^2)$
and if $m_{\phi}$ and $m_{\chi}$ are at very different
scales, say $m_{\chi} \gg m_{\phi}$,
there is no ideal choice of $\mu$ to control the large
logs.  However such terms arise in a mass-independent
renormalization scheme, which is problematic for multi-mass
cases for which there are disparate mass scales, since
accounting for
decoupling effects can not be done.  It is physically
better motivated to use a mass dependent renormalization 
scheme \cite{mdrs},
in which any field with a mass much bigger than others in the system
has its quantum effects suppressed in powers of the light-to-heavy
mass ratio.  Thus in a mass dependent scheme, quantum corrections
lead to terms in the perturbative expansion moderated by power suppression,
$\sim (\mu/m_{\chi})^k \ln^n(m_{\phi}^2/\mu^2) \ln^m(m_{\chi}^2/\mu^2)$.
In this case the choice of renormalization scale 
$\mu^2 \ll m_{\chi}^2$, although will lead to large log terms, it is
not important since the power suppression term dominates the perturbation
series and keeps it under control.

Resummation of the logs is done by applying
renormalization group (RG) techniques to obtain the RG-improved
effective potential \cite{kastening,bando1,ford,chung,einhorn,2loopV}. 
For multimass scale problems,  the prescription given in
Refs. \cite{bando2,casas} is to always choose the renormalization
scale to be of order the lowest mass scale, because
the decoupling theorem will ensure that the heavy mass states do
not contribute. This can be seen directly when using  a mass dependent
renormalization (MDR) scheme, because at any order the logs are modulated by
the appropriate threshold functions \cite{mdrs,nakano}. Recall that these threshold
functions  suppress those massive contributions that are much above the
renormalization scale. Therefore, in the MDR scheme one can immediately see
that one gets the most rapid convergence of the perturbative expansion
by choosing the RG scale below all thresholds. 

Still, the parameters in the effective potential have to be specified
at some renormalization scale, and there is no direct information in
the potential about how to choose that scale. A large value of the
potential, or its curvature, etc..., has no direct relation with the
choice of the renormalization scale. One needs outside information on
the model. For example, when computing the Higgs effective potential
one does not know the self coupling parameter at the electroweak
scale. One procedure then is to invoke a higher symmetry, like a
Grand Unification Theory (GUT) model, which then specifies the
coupling at the GUT scale, and use 
the renormalization group  equations (RGE) to run it down. This is the
standard procedure when studying the effective potential in both the
Standard Model and the Minimal Supersymmetric Standard Model. 
If there is no theoretical argument from
the symmetries of the model, another possibility would be to do some
scattering experiment from which to extract the value of the
self-coupling at some momentum scale, similarly to what is done for gauge
couplings \cite{PDG}. Then, repeating the experiment at different energy
scales one can further confirm the predicted running \cite{PDG}. 

In inflation and quintessence models most often they are not embedded
in a higher theory, and moreover outside phenomenological information
about the parameters is unavailable. For example, in chaotic inflation
normalising the potential to the cosmic microwave
background (CMB) amplitude leads to a tiny
$\phi$ self-coupling of 
$\lambda_{\phi} \simeq 10^{-14}$ \cite{wmap,ci}. This was all
done through a classical calculation. What renormalization scale has this
specification of 
$\lambda_{\phi} \simeq 10^{-14}$ been done at? No information
internal to this calculation tell us about the scale. This is a common
problem in this sort of inflationary model building. As we will
discuss in this paper, this implies an arbitrariness in the model
predictions.

In the case of our model, there are three different couplings
that need to be specified, the
$\phi$ self-coupling $\lambda_{\phi}$, the $\chi$ self-coupling
$\lambda_{\chi}$, and the coupling
$g$ between the two scalar fields. If we could
scatter particles, that would give the couplings and masses at that
scale, but that possibility is not available. 
Indeed, this problem is general to any scalar field inflation
and/or quintessence calculation. One has no {\it a priori} information about
the values of the couplings at any given renormalization scale, so it
is a matter of choice, which leads to considerably ambiguity in the 
predictions from any given model. 
To highlight this point about the ambiguity, decoupling has an
interesting implication. If one chooses all the parameters at a scale 
well below all thresholds in the model, then there will be no quantum
correction to the effective potential as such, and thus the tree-level
potential is the full effective potential. This is the implication of
decoupling in application to the effective potential in
Refs. \cite{bando2,nakano,casas}. On the other hand, if there are
massless states, or one only knows the value of the parameters at some
scale intermediate to the masses in the theory, then one must
implement the renormalization group improved potential.

In our model, we have three couplings $\lambda_{\phi}$,
$\lambda_{\chi}$, and $g$, and the only thing we
know is that either for inflation or quintessence  
$\lambda_{\phi}$ must be
very tiny, although  $\lambda_{\chi}$ and $g$ are 
unconstrained. We also know that the
$\chi$ mass is heavy when the inflaton background field value
is large, whereas the inflaton mass
must be smaller than the Hubble rate, or in the case of quintessence
field, it must be very tiny.
Following the above logic of a scattering experiment,
the energy scale certainly will be below the Planck mass, and an
internal consistency for our model would 
require $\lambda_{\phi}$ to be very
tiny. Then decoupling requires that heavy states yield no radiative
corrections.     

In this paper the decoupling concept will be developed and
applied to inflation and quintessence models.
In section \ref{sect2} we will first discuss quantum corrections due to
renormalizable interactions to other scalar field, computing the
 RG improved effective potential at one-loop
in a MDR scheme. Details about the calculation of
the RGEs in the MDR scheme are provided in Appendix A with one-loop
results, Appendix B with some two-loop results and the one-loop
effective potential is given in Appendix C.
In section \ref{sect3} we discuss the non-renormalizable
self-interactions relevant for a quintessence 
field. In order to check the stability of the potential
against quantum corrections, we study in section \ref{results} 
an example for an inflation
model example, and a quintessence potential. We present the summary
in section \ref{sum}.

\section{One-loop corrections: renormalizable interactions}
\label{sect2}

In order to proceed, we will first check the one-loop corrections due
to the renormalizable interactions of an extra $\chi$ field coupled to
the scalar field $\phi$. We assume that $\phi$ is a light field with a
non-vanishing vev, whereas the $\chi$ state is heavy (heavier than the
Hubble expansion rate), with zero vev. The tree level potential is 
given by: 
\begin{equation}
V^{(0)}(\phi, \chi) =  
\Omega + \frac{1}{2} m^2_{\phi} \phi^2 +\frac{\lambda_\phi}{4!} \phi^4 
                + \frac{1}{2} m_{\chi}^2 \chi^2 
                + \frac{\lambda_{\chi}}{4!} \chi^4   
                + \frac{1}{2} g^2 \chi^2 \phi^2 \,,
\label{Vtree}
\end{equation}
 Loop corrections can give rise to a cosmological constant term
 $\Omega$, a quartic self-interaction $\lambda_\phi$ and mass
 term $m_\phi^2$, and for
consistency we include them already at tree-level. The aim is to show
that those interactions do not pick up large corrections due to the
heavy field, and they can 
remain as small as required  during the inflationary or quintessence
phase. We do not couple the $\phi$ directly to fermions
for simplicity, and focus only on scalar couplings, but we allow a
Yukawa coupling of the $\chi$ field to $N_F$ massless Dirac fermions,
\begin{equation}
\mathcal{L}_{\rm{Yuk}} = -h \chi \bar{\psi} \psi.
\end{equation}
One-loop corrections are given by the Coleman-Weinberg potential
\cite{Veff1}:
\be
\Delta V^{(1)}= \frac{1}{2} \int \frac{d^4 q}{(2 \pi)^4} 
 \sum_\alpha \ln (q^2 + M_\alpha^2) \,, 
\label{V1int}
\ee
where $\alpha= \phi,\, \chi$,  and $M_\alpha$ are the field-dependent masses:
\bea
 M_\chi^2  & =&  g^2 \phi^2 + m_\chi^2 \,, \\  
 M_\phi^2  & =& \frac{\lambda_\phi}{2} \phi^2 + m_\phi^2 \,,
\eea 
(We will denote $\phi$ for both the quantum field as in Eq. (2)
and in all the Lagrangians in this paper as well as for the background
field value as in the above expressions for $M_\chi$ and $M_\phi$,
since the correct usage will be obvious in each case.)
The divergent integrals in Eq.~(\ref{V1int}) can be regularized by
using a cut-off $\Lambda$, and  
keeping only terms that do not vanish when the cut-off goes to
infinity we have:
\be
\Delta V^{(1)}_{reg}= \sum_\alpha \left( \frac{ M_\alpha^2}{32 \pi^2} 
\Lambda^2 + \frac{ M_\alpha^4}{64 \pi^2} \left( \ln
\frac{ M_\alpha^2}{\Lambda^2} - \frac{1}{2} \right) \right) \,,
\label{V1reg}
\ee 
For renormalizable tree-level potentials, the cut-off
divergent terms are subtracted by  adding counterterms:
\be
V_{ct} (\phi)= \delta \Omega + \frac{1}{2} \delta m^2_{\phi} \phi^2
+\frac{\delta \lambda_\phi}{4!} \phi^4 
\ee
and imposing
suitable renormalization conditions at some arbitrary scale $\mu$ on
the effective potential \cite{Veff1}. 
This allows to 
remove the quadratic and logarithmic divergent term, by
choosing:
\bea
\delta \Omega &=& - \frac{\Lambda^2}{32 \pi^2} ( m_\phi^2 +m_\chi^2)  
+\frac{1}{64 \pi^2} ( m_\phi^4 +m_\chi^4)  \left( \ln
\frac{\Lambda^2}{\mu^2} -1) \right) \,, \\
\delta m_\phi^2 &=& - \frac{\Lambda^2}{32 \pi^2} (\lambda_\phi + 2 g^2) 
+\frac{1}{32 \pi^2} ( \lambda_\phi m_\phi^2 + 2 g^2 m_\chi^2)  \left( \ln
\frac{\Lambda^2}{\mu^2} -1) \right) \,, \\
\delta \lambda_\phi &=& \frac{1}{32 \pi^2} ( 3 \lambda_\phi^2 + 12
g^4)  \left( \ln \frac{\Lambda^2}{\mu^2} -1) \right) \,. 
\eea
Adding $V_{ct}$ to Eq. (\ref{V1reg}),
the divergent terms cancel out and one is left with the
logarithmic contributions depending now on the renormalization scale
$\mu$: 
\be
\Delta V^{(1)} = \frac{1}{64 \pi^2} \sum_\alpha M_\alpha^4 \left( \ln \frac{M_\alpha^2}{\mu^2} - \frac{3}{2}\right) \,.
\label{V1ren}
\ee 
Notice however that there is an arbitrariness in choosing the
finite terms in the counterterms, and different renormalization
conditions lead to different finite contributions \cite{ramond} also
in the effective potential. The above prescription has been chosen in 
order to match the standard result for the 1-loop effective potential
using dimensional regularization and minimal subtraction
($\overline {\rm MS}$) as a renormalization prescription. 

Independently of the implemented renormalization scheme, physics
cannot depend on the arbitrary renormalization scale $\mu$, 
and the effective potential $V=V^{(0)} + \Delta V^{(1)}$ has to satisfy
the renormalization group equation (RGE): 
\be
{\cal D} V = \left( \mu \frac{\partial}{\partial \mu} +
\beta_{\lambda_a} \frac{\partial}{\partial \lambda_a} 
- \gamma_\phi \phi \frac{\partial}{\partial \phi}
- \gamma_\chi \chi \frac{\partial}{\partial \chi}
\right) V =0 \,, \label{DV}
\ee
where $\lambda_a$ denotes both renormalizable couplings and mass
parameters in the potential, $\beta_a=\partial \lambda_a/ \partial \ln
\mu$  their beta functions, and $\gamma_\phi$,
$\gamma_\chi$ are the anomalous dimensions of the fields\footnote{The
  anomalous dimensions are the logarithmic $\mu$ derivatives of the
  wave-function renormalization constants of the fields
  $Z_{\phi_\alpha}$. We have included in Eq. (\ref{DV}) the
  contribution of $\gamma_\chi$ for the sake of generality, but we
  only consider situations when $\chi=0$, $\chi$ referring to the vev
  of the field.}. 

 The solution 
to the RGE then provides the RG-improved effective potential
\cite{kastening,bando1,bando2,ford,chung,einhorn,2loopV}, given by:  
\be
V(\phi, \chi, \lambda_a; \mu)= V( \phi(t),
\chi(t), \lambda_a(t);  e^t\mu) \,. 
\label{Vren}
\ee
We can now evaluate the effective potential at any given scale $
t$ by appropriately changing fields, couplings and masses:
$\phi(t)$, $\chi(t)$, $\lambda(t)$ are now 
running parameters, with scale dependence ($t$-dependence)
given by the corresponding 
RGEs.  As remarked in \cite{bando1}, order by order in perturbation
theory, the $Lth$-to-leading log
order RGE improved effective potential is given by  the ($L+1)$-loop  RGE
functions, and the $L$-loop
effective potential at some boundary value of $t$. 
The main idea of the RG improved method is that by choosing 
adequately $t$, the potentially large logs appearing on the LHS of
Eq. (\ref{Vren}) can be resummed. 
Thus, at lowest order the
RG improved effective potential reduces to the tree-level potential
with couplings, masses and fields given by the 1-loop running parameters: 
\bea
V^{eff}&=& \Omega(t)+ \frac{1}{2} m^2_{\phi}(t) \phi^2(t)
+\frac{\lambda_\phi(t)}{4!} \phi^4(t)   \nonumber \\
         &&  \left.     + \frac{1}{2} m_{\chi}^2(t) \chi^2(t) 
         + \frac{\lambda_{\chi}(t)}{4!} \chi^4(t)   
         + \frac{1}{2} g^2(t) \chi^2(t) \phi^2(t)\right |_{t=t_*}\,. 
\label{Veff}
\eea
The key point is how to choose $t_*$, i.e, which is the best choice to
evaluate the effective potential. As stressed in Ref. \cite{casas},
one can choose either a value at which the one-loop potential has the
least $\mu$-dependence, i.e. 
\be
\left. D ( V^{(0)}+ \Delta V^{(1)}) \right |_{(t_*)}=0 \,, 
\ee
or the scale at which the loop expansion has the best apparent
behavior, i.e., $\Delta V^{(1)}(t_*)=0$. The optimal situation occurs
when both criteria are met by the same choice $t_*$. 
This can be done in a model with a single mass scale, say a scalar
field $\phi$ with
self-interactions and no couplings to other fields, so that the choice 
\cite{kastening,bando1,chung,einhorn,2loopV}:
\be
t_* = \ln \frac{\mu_*} {\mu} =\frac{1}{2}\ln \frac{M_\phi^2(t)} {\mu^2} \,,
\ee
fulfills both conditions, and is equivalent to evaluating the one-loop
potential at the scale $\mu= M_\phi$. 
But in the presence of very different mass scales, say $M_\phi \ll
M_\chi$, the choice is not obvious. The problem is how to rearrange
the loop expansion in terms of small parameters for the series
expansion to make sense (and to be resummed). With two different mass
scales, the correction to the effective potential can be
written as \cite{bando2,nakano}:
\be
\Delta V  = g^2 \phi^4 \sum_{l=L-(i+j)}^{\infty}
\left( \frac{g^2}{ 16 \pi^2} \right)^l
\sum_{i,j}^{\infty} F_{i,j}\left[M^2_\chi/M^2_\phi,\lambda_\phi/g^2
  \right] s_\chi^i s_\phi^j \,,
\label{DVsum}
\ee
where 
\be
s_\alpha= \frac{g^2}{16 \pi^2} \ln \frac{M_\alpha^2}{\mu^2} \,.
\ee
By taking as a boundary condition either $s_\phi=0$ or $s_\chi=0$, to
evaluate the $L$-loop effective potential at given order, there are
still potentially large contributions due to $\ln M_\chi^2/M_\phi^2$
in Eq. (\ref{DVsum}), and the series cannot be truncated at any order.    

However, as remarked before, one
does not (and should not) expect heavy states to modify the low energy
physics, and thus the main
issue to address is how to incorporate decoupling of heavy 
states in the improved effective potential \cite{bando2,nakano,casas},
that is, how to get rid of the troublesome large logs by adopting a 
physical condition. 
For example, in Eq. (\ref{Veff}) the decoupling can be incorporated in
the running parameters through their RGEs. In a mass-independent
renormalization scheme, with no reference to mass scales, decoupling
has to be implement by hand by the use of step functions in the RGEs
and matching conditions for masses and couplings at each
threshold: for a given state, when its mass becomes heavier than the
renormalization scale, its contribution drops from the RGEs and the
effective potential. In particular we will have \cite{casas}:
\bea
\Delta V^{(1)} &=& \sum_\alpha \theta(\mu^2 - M_\alpha^2)
V^{(1)}_\alpha \,, \label{Vcasas}\\
V^{(1)}_\alpha &=& \frac{M^4_\alpha}{64 \pi^2} \left(\ln
\frac{M_\alpha^2}{\mu^2} -\frac{3}{2} \right) \,, 
\eea
and  the optimal choice
for $t_*$ (or $\mu_*$) is then given by the lower threshold in the
model. Indeed in that case all massive states are decoupled, so that 
$\Delta V^{(1)}(\mu_*)=0$, i.e., the effective potential
is given by the tree-level potential with masses and couplings
evaluated at low energy. The effect of the heavy states appear when
integrating the RGEs from high energy down to the low energy regime. 

However, threshold
effects are more naturally taken into account when adopting instead a
mass dependent renormalization (MDR) scheme \cite{nakano}. Following
Ref. \cite{mdrs}, effective couplings and masses are defined after
subtracting the divergences (regularized using dimensional
regularization for example) of the 1PI Green functions, by imposing
suitable normalization conditions at the euclidean external 
momentum $p^2_E= -p^2=\mu^2$, with $\mu^2$ being the arbitrary
renormalization scale. The Appendices give a detailed account of the
MDR scheme, beta 
functions and 1-loop correction to the effective potential  
in this scheme, while the following just summarizes the
main results.

We have already mentioned the scheme-dependence of the finite
contributions in the counterterms, or equivalently the renormalization
constants $Z_a$ \cite{ramond}:  
\bea
\delta m_\alpha^2 &=& m_\alpha^2 (Z_{m_\alpha^2}^{-1}-1 ) \,, \label{deltam}\\
\delta \lambda_\alpha  &=& \lambda_\alpha (Z_{\lambda_\alpha}-1) \,, \label{deltalambda}\\
\delta \Omega &=& \Omega (Z_\Omega^{-1}-1) \,, \label{deltaL} 
\eea
that relate the bare parameters (denote by a subindex ``0'') with the
renormalized ones: 
\bea 
m_{\alpha}^2 (\mu)&=& Z_\phi Z_{m_\alpha^2} m_{\alpha 0}^2 \,, \\
\lambda_{\alpha}(\mu)  &=& Z_\phi^2 Z_{\lambda_\alpha}^{-1}
\lambda_{\alpha 0}  \,, \\
\Omega (\mu) &=& Z_\Omega \Omega_0 \,. 
\eea
By modifying the subtraction conditions, we are 
explicitly including finite contributions from the 1PI functions,
which carry the dependence on the 
different mass scales of the model. When taking the derivative with
respect to the arbitrary renormalization scale $\mu$, that dependence
appears in the beta functions as threshold functions depending on the
different ratios $M_\alpha^2/ \mu^2$. These threshold functions
modulate the contribution of each massive state to the running of the
different parameters. And in the exact decoupling limit $\mu=0$ the
threshold functions
vanish. For the renormalizable
parameters in Eq. (\ref{Vtree}), the beta functions for the couplings
and mass parameters are given by:      
\bea
(4 \pi)^2 \beta_{\lambda_\phi} &=& 3 \lambda_\phi^2 F_2(a_\phi) + 12 g^4
F_2(a_\chi) \,, \label{betalam}\\ 
(4 \pi)^2 \beta_{\lambda_\chi} &=& 3 \lambda_\chi^2 F_2(a_\chi) + 12 g^4
F_2(a_\phi) +(8 \lambda_\chi h^2 - 48 h^4) N_F \,, \label{betalamchi}\\
(4 \pi)^2 \beta_{g^2} &=& g^2 (\lambda_\phi F_2(a_\phi) + \lambda_\chi
F_2(a_\chi) + 8 g^2 F_1(a_\chi) +4 N_F h^2) \,, \label{betag}\\
(4 \pi)^2 \beta_{h^2} &=& h^4 (6 F_3(a_\chi) +4 N_F )  \,,  \label{betah}\\ 
(4 \pi)^2 \beta_{m_\chi^2} &=& 4N_Fh^2 m^2_\chi \,, \label{betamphi}\\ 
(4 \pi)^2 \beta_{m_\phi^2} &=& 0\,, \label{betam} \\
(4 \pi)^2 \beta_{\Omega} &=& 0\,, \label{betaL}
\eea
where $N_F$ is the number of massless fermions, and
we have defined $a_\alpha=M^2_\alpha/ \mu^2$. Notice that by using a
MDR scheme,
with no couplings of the field $\phi$ to fermions, the mass parameter
$m_\phi^2$ and the vacuum energy contribution $\Omega$ do not
run: they are fixed by the boundary conditions. That is,
the pure (quadratically) divergent terms from vacuum diagrams with no
reference to the external energy scale can be subtracted from the bare
parameters in the potential, leaving a fixed, finite contribution. For
example we can always impose as normalization condition for the vacuum
contribution: 
\be
\Omega(\mu)=0 \,.
\ee
On the other hand the effective field dependent mass $M_\phi^2$ does
run, due to the running of the coupling constant $\lambda_\phi$. 

\begin{table}
\begin{tabular}{|c|c|c|c|}
\hline
&$c_0$ &$c_1$ & $c_2$\\
\hline
$F_1(a)$ &   15.3& 20.1 &  31.8 \\
$F_2(a)$ &   48.1& 60.04& 295.4 \\
$F_3(a)$ &   48.1& 60.04& 295.4 \\
\hline
\end{tabular}
\label{table1}
\end{table}

\begin{figure}
\includegraphics[width=0.5\textwidth]{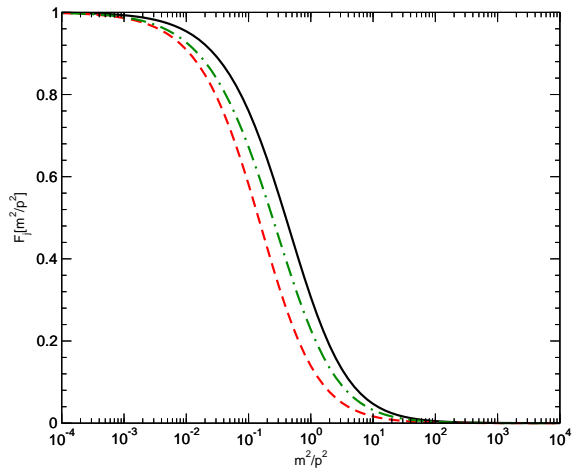}
\caption{\label{plot1} Threshold functions: $F_1(m^2/p^2)$ (solid,
  black), $F_2(m^2/p^2)$ (dashed, red),  and $F_3(m^2/p^2)$
  (dot-dashed, green).  }
\end{figure}

The expressions for the threshold functions $F_i$, $i=1,2,3$, are given in
Eqs. (\ref{F1}) - (\ref{G1}) in Appendix A (also some two-loop
expressions are given in Appendix B).   
They can  be well approximated by:
\be 
F_j(a) \simeq \frac{1 + c_{0j}
  a}{1 + c_{1j} a+ c_{2j} a^2} \,, 
\ee 
with the coefficients $c_{ij}$ for each function given in Table \ref{table1}. 
In the massless limit, all threshold functions reduce to one, which
recovers for the couplings (dimensionless parameters)  the standard
RGEs computed for example in $ \overline {\rm MS}$ (see Appendix A). On
the other hand, when the ratio $a_\alpha \gg 1$, we have $F_j(a_\alpha)
\simeq O(1/a_\alpha)$, i.e., power suppression of the heavy state
contribution.  Decoupling is not instantaneous, as can be seen in
Fig. \ref{plot1}. Threshold functions 
smoothly interpolate between the high energy regime where massive
states can be viewed as massless, and the low energy theory without
heavy states.  

All that remains now is to fix the initial conditions at some scale
$\mu$ to integrate the RGEs and obtain the values of parameters when all
masses decouple. With those we evaluate the tree-level potential 
to obtain the RG-improved effective potential at 1-loop. 
It is shown in Appendix C that by substituting back the solution for
the running couplings at low energy in the tree-level potential one
recovers the one-loop correction computed in the MDR
scheme:
\bea
\Delta V^{(1)} &=& \frac{1}{64 \pi^2} \left( 
\frac{\lambda_\phi^2 \phi^4}{4}( \ln \frac{M_\phi^2}{
  \mu^2} -I(M_\phi^2/ \mu^2))       
+ g^4 \phi^4( \ln \frac{M_\chi^2}{
  \mu^2} -I(M_\chi^2/ \mu^2)) \right) \,, 
\label{V1md}
\eea       
where
\be
I(a) = \ln a -2 - \sqrt{1 + 4 a} \ln
\frac{\sqrt{1+ 4 a}-1}{\sqrt{1+ 4 a}+1} \,.
\ee
The main difference between the one-loop correction computed in a mass
independent renormalization procedure, and the MDR scheme one, comes into the
non-logarithmic contribution, and it is due to the different
scheme-dependent finite contributions in the renormalization
conditions. Comparing Eq. (\ref{V1md}) with Eq. (\ref{V1ren}), 
the constant ``3/2'' term is replaced by a threshold function
$I(a)$, which controls the contribution of the original log
term. Thus, whatever the hierarchy among the masses, we obtain the exact
result at 1-loop:
\be
\Delta V^{(1)}(\mu=0) = 0 \,.
\ee  
In section \ref{results} we will present some examples of the procedure for an
inflation model, and a quintessence one. We want to check the
impact of the radiative corrections on the inflaton/quintessence
potential as $\phi$ changes. Notice that by changing $\phi$ we are
implicitly changing the threshold conditions that depend on the
effective masses, and therefore the values of the couplings, so
effectively what we have are field dependent couplings. We
will check that at least during the regimes of interest, when
$\phi$ evolves the couplings remain in the perturbative regime and
none of them picks up a large correction. For example for the
quintessence model, we can impose that we have indeed a quintessence
regime, such that for $\phi \gg m_P$ the quartic coupling
$\lambda_\phi$ is tiny (or even cero). By going backwards in time,
i.e., taking smaller values of the field, the evolution should be such
that the coupling never gets large
enough to disturb the standard evolution of the quintessence field.

\section{One-loop corrections: non-renormalizable self-interactions}
\label{sect3}

Most Quintessence potentials (and some inflation potentials)
are typically given by non-renormalizable
$\phi$ potentials, $V_{NR}(\phi)$, so that the tree-level potential is now
given by:
\begin{equation}
V^{(0)}(\phi, \chi) = V_{NR}(\phi)+ \frac{1}{2} m^2_{\phi} \phi^2 +\frac{\lambda_\phi}{4!} \phi^4 
                + \frac{1}{2} m_{\chi}^2 \chi^2 
                + \frac{\lambda_{\chi}}{4!} \chi^4   
                + \frac{1}{2} g^2 \chi^2 \phi^2 \,,
\label{VtreeNR}
\end{equation}
The $\phi$-dependent mass of the $\phi$ field picks up an extra term due to the
non-renormalizable interaction, and in Eq. (\ref{V1reg}) we have to
replace $M_\phi^2$ by $V_{NR}^{\prime \prime} + M_\phi^2$.  
The 1-loop correction $\Delta V^{(1)}$ can be split into a
renormalizable and a non-renormalizable contribution, owing to the
origin of the field-dependent masses:  
\be
\Delta V^{(1)}= \Delta V^{(1)}_{NR}+ \Delta V^{(1)}_{\rm ren} \,,
\ee
where $\Delta V^{(1)}_{\rm ren}$ is given in Eq. (\ref{V1reg}), with
$M_\alpha= M_\phi,\,M_\chi$, while $\Delta V_{NR}$ is given by: 
\bea
\Delta V^{(1)}_{NR}&=& \frac{V_{NR}^{\prime \prime}}{32 \pi^2} 
\Lambda^2 + \frac{1}{64 \pi^2} \left( V^{\prime \prime}_{NR} ( V^{\prime
  \prime}_{NR} + 2 M_\phi^2) \left( \ln
\frac{ V^{\prime \prime}_{NR} + M_\phi^2}{\Lambda^2} - \frac{1}{2} \right)  
+ M_\phi^4 \ln \frac{ V^{\prime \prime}_{NR} + M_\phi^2}{M_\phi^2}  \right)
\,. 
\eea
Having dealt with the renormalizable interactions in the previous
section, we come back to  the quadratic cut-off and log 
dependent term due to the non-renormalizable interaction. The standard
approach, which we will 
follow, is then to consider $\Lambda$ as the effective ultraviolet
cutoff for the model, such that:
\bea
\Delta V^{(1)}_{NR} &\simeq& \frac{V^{\prime \prime}_{NR}}{32 \pi^2}
\Lambda^2 \,, 
\eea
where in $\Delta V^{(1)}_{NR}$ we have kept only the dominant quadratic
contribution \cite{Doran:2002qd}.   

Without loss of generality, let us consider a generic quintessence potential
of the form:  
\be
V_{NR}\left(\phi \right) =  \displaystyle{\frac{\lambda M^{(4+n)}}{\phi^n +
    M^n}}  \,. 
\label{Vquint}
\ee
The scale $M$ is
model dependent, and we do not consider any particular value; here it only
parametrizes the value of the field well outside the quintessence
phase, $\phi \sim M$, while the  quintessence regime happens for
values of the field $\phi \gg m_P$.  Now we want to check that indeed
$\Delta V^{(1)}_{NR}$ does not provide large corrections to the
potential, $\Delta V^{(1)}_{NR} \ll V_{NR}$. This condition is not
difficult to fulfill when $\phi \gg 
M$, and in this regime we simple have $\Delta V_{NR}/V_{NR} \simeq
(\Lambda/\phi)^2$,  
so in the quintessence regime when $\phi > m_P$, 
the mass squared $V_{NR}^{\prime \prime}$ has become tiny, and the
effective ultraviolet cut-off can be taken 
close to the Planck scale and still $\Lambda/\phi < 1$. 
However, at early times when $\phi \sim M
\ll m_P$ we have that $\Delta V^{(1)}_{NR}/V_{NR} \simeq (\Lambda/M)^2$, which
can be large unless the cut-off is well below $M$. 

Nonetheless, one
can argue as done in Ref. \cite{Brax:1999yv} that this is all right as
far as the one-loop contribution is suppressed not with respect to the
tree-level potential, but with respect to the dominant energy density
at the time. For example there are some restrictions on the amount of
dark energy at the time of BBN, with $\rho_{DE}< 0.2 \rho_{\rm rad}$, 
which in this case should be satisfied by the 1-loop effective
potential. At earlier times, the quintessence field may find itself
fast rolling the potential, with the Universe 
dominated by its kinetic energy density $\rho_{KE}$ (kination). 
The condition to be on the safe side would be that still $\Delta
V^{(1)}_{NR} \ll \rho_{KE}$. 
However, it is not clear how to reconcile
a fast rolling field with the calculation of the improved effective potential,
and the approximation may break down. 
Because of that we do not pursue the calculation of the effective
potential into that regime. Whenever a kination phase due to the
quintessence field in the early universe, we can check that after that
quantum corrections  do not mess up the evolution of the quintessence
field, and that quintessence domination is reached today. Going
backwards in time, if the quantum corrections are subdominant by the
time of kination, we assume that they will not grow as much as to
change this phase. We have no means to consistently check this
assumption, but we consider it a reasonable working hypothesis. 

\section{Results for inflation and quintessence} 
\label{results}

\subsection{Inflation}
Let us consider inflationary potentials of the form: 
\be
V\left(\phi \right) =  \lambda_\phi M^4\left(\frac{\phi}{M}\right)^n\,,
\ee
which is added to $V(\chi)$ in Eq. (\ref{v2f}).
Due to the coupling between $\phi$
and $\chi$, radiative corrections will always induce a quartic
interaction for the inflaton field. Just to keep the discussion
simple, we focus on the $n=4$ case. The tree-level potential
reduces then to Eq. (\ref{Vtree}), with\footnote{We have shown in
section \ref{sect2}
that the mass parameter $m_\phi$ does not run in the
MDR scheme, so that imposing as a boundary condition $m_\phi(\mu)=0$
ensures that this parameter vanishes at 1-loop at any other energy
scale.} $m_\phi=0$  

In such chaotic potentials inflation takes place for $\phi > m_P$, and
therefore the $\chi$ field gets a large mass $g \phi \simeq m_P$, for
moderate values of the coupling $g$. On the other hand, the inflaton
mass is $M_\phi^2 = V_{\phi \phi} < H^2 \ll M_\chi^2$. Then, following
our prescription, the appropriate renormalization scale $\mu=\mu*$
for examining physics during inflation is of
the order of $M_\phi$. At this scale, the threshold functions imply
a suppression of the effect of the $\chi$ loops in the renormalization
group equations. 



Notice that in the standard $\overline{\rm MS}$ scheme, the generic
approach is to suppress large logs, in which case at one-loop
order it would mean taking
$\mu* \simeq M_\chi$. However, in the MDR scheme the large log
contribution is suppressed by the threshold function prefactor when
$\mu$ is chosen at the lowest threshold. In fact, at two-loops at
order $g^4$, there   
will be large logs depending on both masses $M_\phi$ and $M_\chi$, and
in the $\overline {\rm MS}$ scheme it is not clear which is the best choice for
$\mu*$. Nevertheless, the MDR scheme is unambiguous that it is
around the lowest mass scale. 

Up to order of magnitude, the above approach fixes the choice of $\mu*$
in the MDR scheme. But
there is an uncertainty in the exact value one should choose. This
underlies an inherent ambiguity in the value of the effective
potential, which ultimately implies a theoretical uncertainty in the
coupling and thus on the model predictions such as the amplitude of
primordial perturbations.  If the effective potential could be
computed exactly, then it would be completely $\mu$-independent
and any choice of $\mu*$ would be equally good.  However in
any real calculation, where the effective potential is
calculated only to some finite order, often just one-loop order,
the choice of $\mu*$ must be made carefully.  For a given
choice of $\mu*$,
slightly larger or smaller values should
result in the same answer, and if they do not, the selection of $\mu*$ is
flawed for the given order in the loop expansion.


To implement the calculation of the RG improved effective potential,
two renormalization scales in general are needed. First is the scale,
which will be called ${\bar \mu}$, where the initial values of the
parameters in the theory are specified and second is the scale where
we want to use the potential to do physics, which we have already
denoted $\mu*$.  A further detail in specifying the initial values of
the parameters at ${\bar \mu}$ is this specification in general must
be given over some range of $\phi$, and thus the masses in the system.
As $\phi$ changes, these $\phi$-dependent masses will change, thus
this choice of ${\bar \mu}$ and/or initial values of the system
parameters can change.

There are two approaches we will consider for initializing the RG improved
calculation of the effective potential, which will be referred to as the
high and low energy approaches.  In the low-energy approach the parameters
of the system are initialized at the scale where one is interested
in using the effective potential to do physics, thus
${\bar \mu} = \mu*$.  Since this $\mu*$ in MDR scheme is at the scale
of the smallest mass, it means there would be very small
quantum corrections to the effective potential from heavy mass states.
Moreover as $\phi$ changes, thus the masses in the system, 
in principle one could
use RG to move the value of $\mu*$ to optimize the quantum corrections,
although the effects would be small.
In the high energy approach,
the values of the initial parameters are specified at some high
renormalization scale ${\bar \mu}$ and then the renormalization
group is used to run the parameters down to the scale $\mu*$,
where physics is to be done.

An example of implementing the low-energy approach in the
case of inflation would be to fix the value of
$\lambda_\phi$
at the epoch of inflation corresponding today
to the largest observable scale. This value of 
$\lambda_{\phi}$ can be determined from density perturbation
constraints from measurement of the cosmic microwave background.
The renormalisation scale this corresponds to
would be (following our prescription) the lowest mass scale in the
theory, $\mu* \simeq M_\phi$. We may therefore say that at these values
of $\phi$ and $\mu$, observations tell us $\lambda_\phi(\mu*) \approx
10^{-14}$, and this is all. If we have a more complicated model with
additional parameters, we must also be able to specify the remaining
parameters such as $g^2$ also at $\mu*$.
In this approach there would be no or very small
quantum corrections, so that the tree-level potential would
be almost identical to the RG-improved one. 

The high energy approach might be implemented if the scalar potential
were embedded in a higher theory, and some symmetry at a
high energy scale ${\bar \mu}$ specified the value of the parameters and
over some range of $\phi$.  Then the RGE could be used
to run the parameters down from $\bar \mu$ to $\mu*$ where one
wishes to do physics with the effective potential.

An important point here is the matter of initial conditions is not
simply a mathematical concern.  There is in general missing
\emph{physics} in inflationary models. The predictions one obtains
from such models depend on this missing physics.  Thus for a given
inflationary potential, depending what higher theory it is embedded
in, different specifications might emerge for the value of the
coupling $\lambda_{\phi}$ at some high energy scale, which when
evolved down to $\mu*$ will lead to different predictions for
inflation and large scale structure. This can be viewed in an
alternative fashion. In the low energy approach, we require the
inflation model to be consistent with observation, thus the parameters
are determined by the physics of inflation.  This might then be used
to place constraints on the unspecified physics at higher energies.

Let us now examine the behavior of the parameters with RG running.
For the high energy approach at some renormalisation scale $\mu =
\bar{\mu}$ some unknown physics specifies that the parameters (and
$\lambda_\phi$ in particular) maintain their values over a range in
$\phi$. Although we are not considering any specific model for
the high scale physics, we would like to investigate the procedure
for how this would in principle be done.  
As such, what we will do to determine
the values of the parameters at scale
$\bar{\mu}$ is run from $\mu*$ (where
the parameters are specified or known) to $\bar{\mu}$, at the value of
$\phi$ where a constraint on the parameters exists. For our purposes
this will be $\phi = m_p$. This is just to ensure that the effective
potential will match observed constraints at $\phi = m_p$. We would like to
check it continues to do so at larger/smaller values of $\phi$.

\begin{figure}
\includegraphics[width=0.5\textwidth]{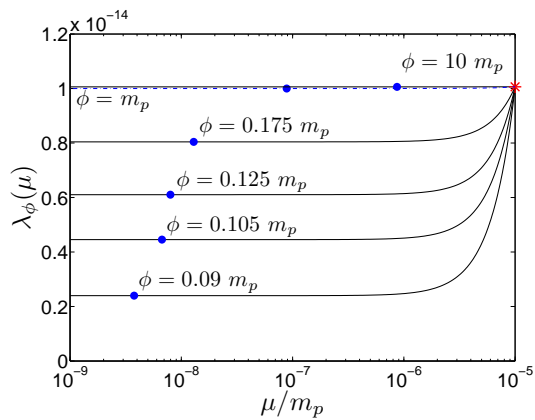}
\caption{\label{plot2}The running of $\lambda_\phi$, with
  $g^2= 10^{-4}$. The red star indicates $\mu=\bar{\mu}$; at
  this scale some assumed physics informs us that $\lambda_\phi$ is
  constant over the considered range in $\phi$. Blue dots indicate the
  position on each curve where $\mu = \mu*$, and where the parameter
  should be taken to improve the effective potential. Curves with
  smaller $\phi$ than those shown quickly drive $\lambda_\phi$
  negative before $\mu*$ is reached. At larger $\phi$ the curves are
  flat and constant, with a value that increases only slightly as
  $\phi$ increases.}
\end{figure}

Thus we start with the following parameters at $\phi = m_p$ and 
$\mu=\mu* = M_{\phi}$:
$\lambda_{\phi} = 10^{-14}$, $g^2 = 10^{-4}$,
$\lambda_{\chi}=10^{-3}$, $h^2 =10^{-4}$, $m^2_{\chi} = 10^{-9}
m^2_P$, $m^2_{\phi} = 0$, $N_F=8$, and $\bar{\mu} = 10^{-5} m_P$. We
run the parameters upwards to $\bar{\mu}$ and find the value of the
parameters there. At this scale our assumed physics keeps these
parameters unchanged with $\phi$. Thus we now change $\phi$, and run back
down to $\mu=\mu*$. This lets us probe the effective potential at 
different values of $\phi$.

What is found is
most parameters remain nearly constant (varying less
than 0.1\% over the range in $\mu$ considered); the only parameter
that runs appreciably is $\lambda_{\phi}$, and is plotted on
Fig. \ref{plot2}. The blue dashed line on Fig. \ref{plot2} shows this
initial curve ($\phi=m_p$) for $\lambda_{\phi}$. The red star denotes
$\bar{\mu}$, which is at the same location on all curves.

Recall at this scale $\bar{\mu}$ the parameters are \emph{assumed} to
be independent of $\phi$: through this assumption we are including the
missing physics. We run down from the red star to the low scale
$\mu*$ for different values of $\phi$ (the blue dots denote the
point $\mu*$ on each line). At this scale, the parameter should be
taken and inserted into the tree-level potential to generate the
one-loop improved effective potential.

It is clear the curves are very flat at the scale $\mu*$; it makes
little difference if the parameters are evaluated at precisely $\mu*$
or within some order of magnitude of this scale (or indeed, many
orders of magnitude below it). This is precisely as we would expect,
as we have chosen $\mu*$ just for this property, so as to satisfy the
$\mu$-independence of our improved effective potential.

\begin{figure}
\includegraphics[width=0.5\textwidth]{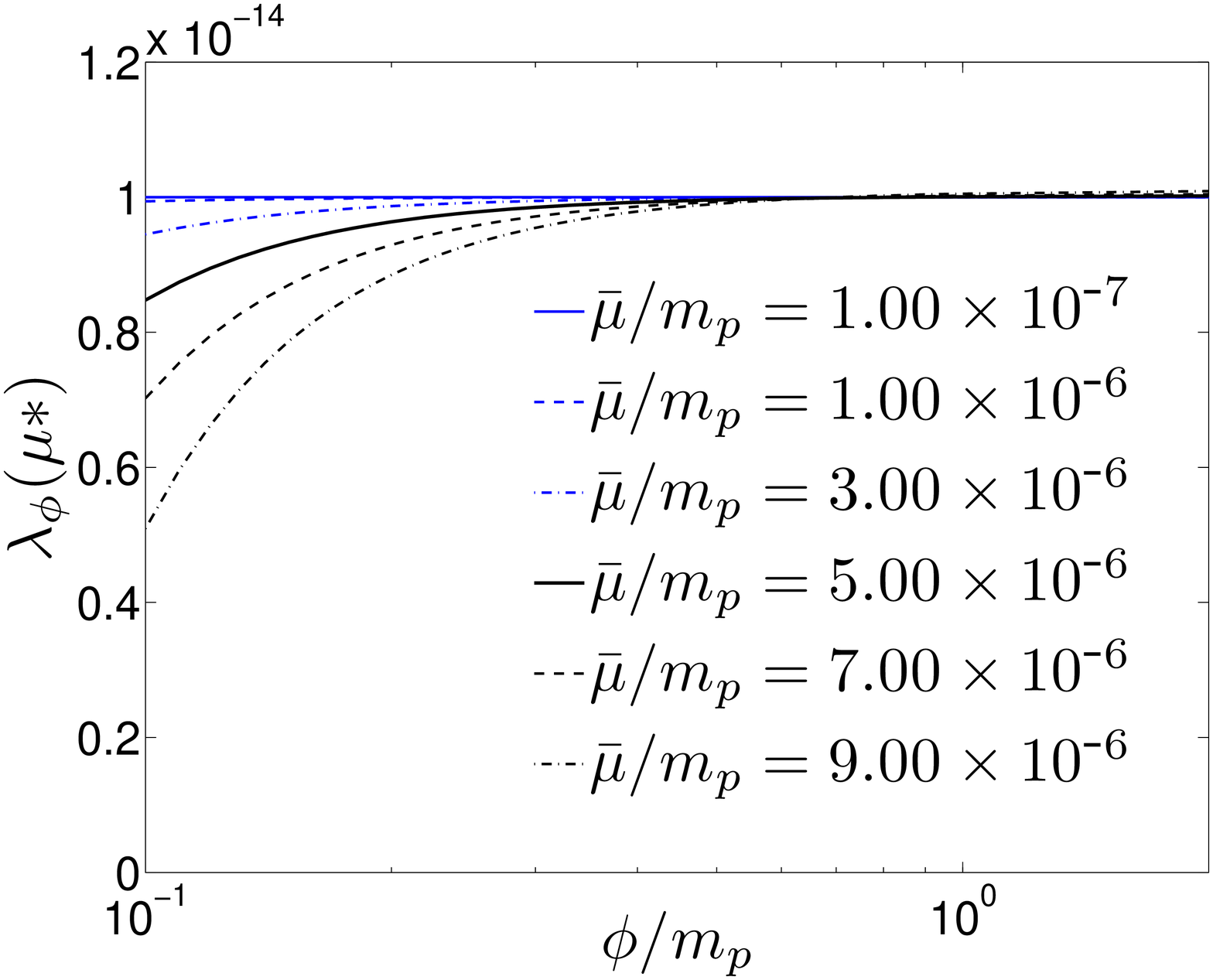}
\caption{\label{plot3} Curves showing how $\lambda_\phi(\mu*)$ varies
with $\phi$, for different values of $\bar{\mu}$. As before, $g^2 =
10^{-4}$. Since $\lambda_\phi(\mu*)$ appears in the RG-improved
effective potential, this is the physically important
quantity. Curves with $\bar{\mu}$ larger than those shown quickly
drive $\lambda_\phi(\mu*)$ negative before $\phi$ is as small as
$0.1 m_p$, and have mildly larger values at large $\phi$.
At larger $\phi$ the curves asymptote to a constant value. Increasing
$g^2$ has a similar effect to increasing $\bar{\mu}^2$, and the
increase is one may be compensated for by a decrease in the other,
allowing for larger values of $g^2$ than plotted here.}
\end{figure}

As each curve generates a single value of $\lambda(\mu*)$ for a single
value of $\phi$, we can construct curves of how $\lambda(\mu*)$ varies
over $\phi$ directly. To do so we compare the values at $\mu*$ (the
blue dots) for a large number of curves. Furthermore, we can examine
how these curves change if the value of $\bar{\mu}$ (the horizontal
position of the red star) is changed. The result is Fig. \ref{plot3},
maintaining the above parameters but varying $\bar{\mu}$.

How can we understand this behaviour? Consider the equation governing
the running of $\lambda_\phi$, Eq. (\ref{betalam}), \be (4 \pi)^2
\beta_{\lambda_\phi} = 3 \lambda_\phi^2 F_2(a_\phi) + 12 g^4
F_2(a_\chi) \,.  \ee We will now make some approximations. If we
assume $\mu^2 \ll M^2_\chi$ (certainly true for the numerical range
shown above), we may use the approximate behaviour of the threshold
function, $F_2(a_\chi) \approx \mu^2/(g^2\phi^2)$. Despite this
suppression, the value of $\lambda_\phi$ is still small enough
for the second driving term to dominate and we find: 
\be (4 \pi)^2
\frac{\ud \lambda_\phi}{\ud \mu} \simeq  12\frac{g^2}{\phi^2} \mu \,.  
\ee
We can solve this if we make the assumption $g^2$ is constant, which
is well supported numerically. The solution is

\be \lambda_\phi(\mu) =
\lambda_\phi(\bar{\mu}) - 
\frac{6g^2}{(4\pi)^2} \left(\frac{\bar{\mu}}{\phi}\right)^2 
 \left[1- \left(\frac{\mu}{\bar{\mu}} \right)^2 \right] \,.  
\ee We see now precisely the
behaviour in Fig. (\ref{plot2}); starting from an initial value at
$\lambda_\phi(\bar{\mu})$, $\lambda_\phi$ quickly reaches a constant
value when $\mu \ll \bar{\mu}$. We also see the $\phi$ dependence
exhibited in Fig. (\ref{plot3}). As $\phi$ is steadily reduced, the
second term becomes larger, eventually dominating the first term and
driving $\lambda_\phi$ negative. Larger values of $\bar{\mu}$ have the
same effect as this.  For $\lambda_\phi(\mu*)$ to be stable against
changes in $\phi$, we require the second term smaller than the
first. This gives the constraint: 
\be 
\frac{6 g^2}{\left(4\pi\right)^2}
  \left( \frac{\bar{\mu}}{\phi} \right)^2 \ll \lambda_\phi(\bar{\mu}) \,,
\ee
which can be written as:
\be
 g \bar \mu \gg 10 M_\phi (\bar \mu) \,.
\ee
For inflation, with $\lambda_\phi(\bar{\mu}) \approx 10^{-14}$ and
at its smallest $\phi = 0.1 m_p$, we find:
\be
g \bar{\mu} \ll 10^{12}\, \textrm{GeV} \,.
\ee
This agrees well with Fig. \ref{plot3}; if $g^2 \approx 10^{-4}$,
then $\bar{\mu}$ may be no larger than $10^{14}$ GeV.

When inflation ends at around $\phi \sim m_p$, the suppression is
reduced compared to earlier, larger field values.  As the field
strength continues to decrease, the effects become gradually
larger. The values of $\bar{\mu}$ and $g^2$ in the plots have been
chosen to display this behaviour between $\phi = m_p$ and $\phi = 0.1
m_p$.

For completeness, note that if a large mass parameter is given to
$\chi$, such that $M_\chi \approx m_\chi \gg g^2 \phi^2$, the solution
becomes independent of $\phi$ at these field values. However, for
the solution to be valid, we would require 
$M_\chi \approx m_\chi \gg \bar{\mu}$ to suppress the $\chi$ loops.

Let us now pause and consider the physical interpretation of these
results. Without decoupling, we would expect corrections to
$\lambda_\phi$ of size $\mathcal{O}(g^4)$. Instead, due to the
suppression of the $\chi$-loops, we find this result is reduced by
a factor of $\bar{\mu}^2/M^2_\chi$.  
Deriving an effective field theory, by integrating out the heavy
degrees of freedom from a more fundamental theory, typically leaves
behind a low energy theory along with terms in the potential
suppressed by factors of $\mathcal{O}\left(E^2/m^2\right)$, where $E$
is the low energy scale the theory is probed at, while $m$ is the
energy scale associated with the heavy degrees of freedom. For
instance, classical electromagnetism encounters corrections with a
size $\mathcal{O}\left(E^2/m^2_e\right)$ where $m_e$ is the mass of
the electron.  We might have expected in the case of inflation to find
quantum corrections suppressed in a similar fashion, with corrections
$\mathcal{O}\left(E^2/M^2_\chi\right)$. But when dealing with an
effective potential directly, it is not clear what energy scale to
associate to $E$. The effective potential is computed from diagrams
with external legs set to zero momentum, and the only other scale
is the renormalisation scale, upon which our results cannot
depend.  We now have an answer (at least for the form we have assumed
our missing physics take): it is the scale associated with
$\bar{\mu}$, where the theory is set. If the extremely large
field-strength associated with the inflaton (with $\phi > m_p$ in
chaotic inflation) serve to generate field dependent mass of the
$\chi$-field far above this scale, its contribution to the effective
potential is suppressed.

At what scale should we expect $\bar{\mu}$ to be? Inflation is
associated with high vacuum energy densities. But the relationship between 
the renormalisation scale and physical energy scale is not
clear. Without an obvious way of relating these two properties, it is
difficult to motivate any particular choice of $\bar{\mu}$ above any
other. 

\subsection{Quintessence}

Now we revert to the generic quintessence potential introduced in
Eq. (\ref{Vquint}).  Quintessence occurs when $\phi \gg m_P$, and
thus, any field at least moderately coupled to $\phi$ will acquire a
large mass from the $\phi$ background field value. Therefore, again one expects
only power corrections to the effective potential from such heavy
states. From the previous example, we can perhaps already guess that
we should expect the typical $g^4$ corrections to
$\lambda_\phi$ to be suppressed by $\mathcal{O}(\bar{\mu}^2/M^2_{\chi})$.


In a similar fashion to inflation, we have only limited knowledge of
the effective potential. We know only that today, with $\phi \simeq
m_p$, the value of $\lambda_\phi$ appearing in the effective potential
must be small enough so that the dominant contribution comes from the
non-renormalizable term $V_{NR}(\phi)$. Another way of stating this is
that the tiny effective field mass of $M_\phi \sim 10^{-33}$ eV
remains unchanged by the size of $\lambda_\phi(\mu*)$. From this, we
can place a constraint on $\lambda_\phi(\mu*) \ll 10^{-124}$.

At earlier epochs in the history of the universe, the quintessence
field strength is smaller. As discussed earlier, at very small field 
strengths the effective potential description breaks down, but we
would like to make sure the potential is not disrupted by quantum
corrections over at least an order of magnitude in $\phi$. 

\begin{figure}
\includegraphics[width=0.5\textwidth]{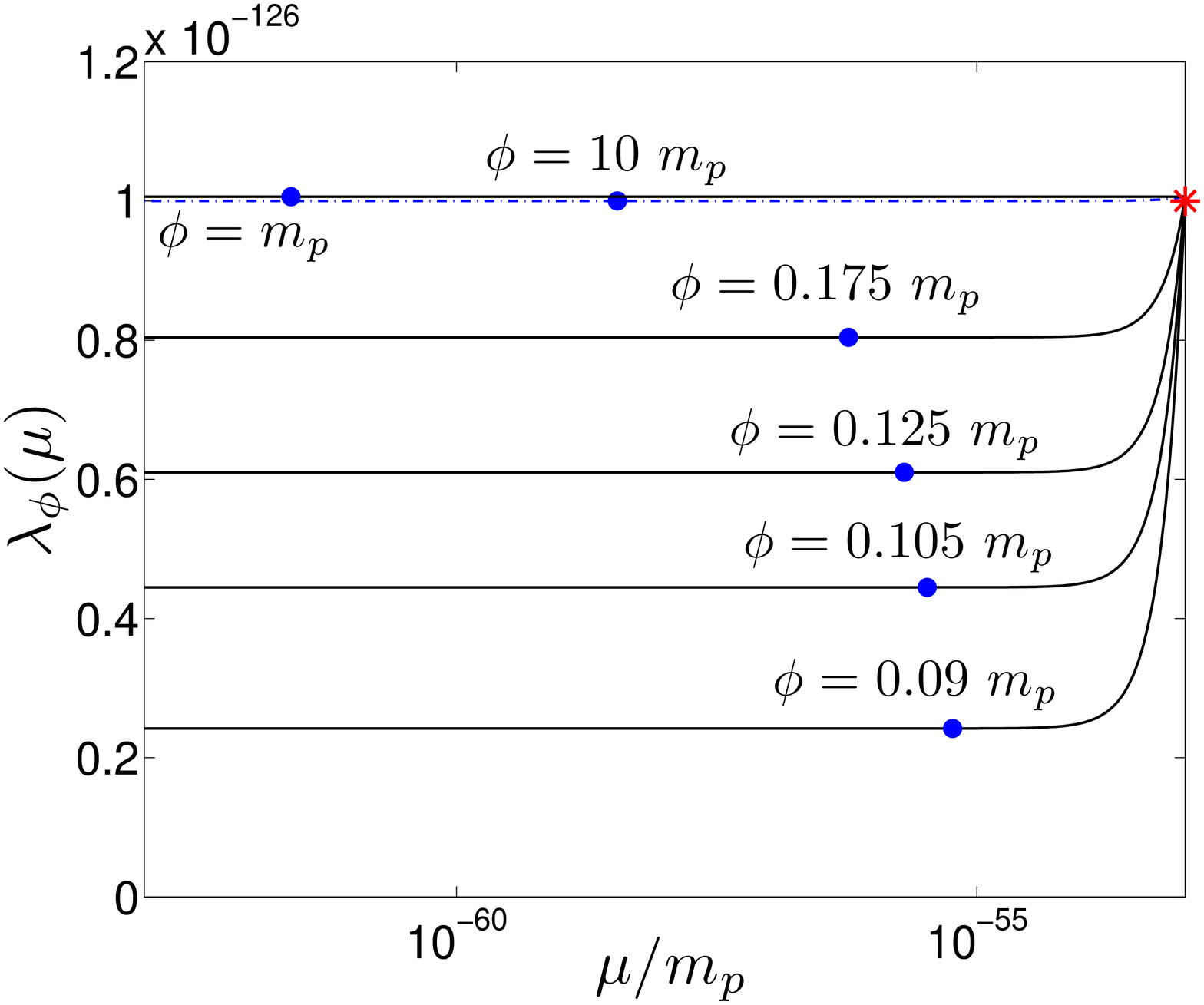}
\caption{\label{plot4}The running of $\lambda_\phi$, with
$g^2 = 10^{-20}$. As with inflation, the red star indicates $\bar{\mu}$ 
where the parameters are assumed to be $\phi$ independent. The
blue dots on each curve indicate $\mu*$ for each value of $\phi$.
Increasing $g^2$ has a similar effect to decreasing $\phi^2$. }
\end{figure}

We proceed in just the same way as in the previous section with
inflation. The system of equations are initialised at $\mu=\mu*$ with
$\phi = m_p$. We take parameters at this scale as: $\lambda_{\phi} =
10^{-126}$, $g^2 = 10^{-20}$, $\lambda_{\chi}=10^{-3}$, $h^2
=10^{-4}$, $m^2_{\chi} = 10^{-9} m^2_p$, $m^2_{\phi} = 0$, $N_F=8$,
and $\bar{\mu} = 10^{-57} m_p$. All is identical as with inflation,
except for the size of $\lambda_\phi$, $g^2$ and $\bar{\mu}$, and of
course the non-renormalizable contribution to $M_\phi$. The result is
Fig. \ref{plot4}, where once again the blue dashed line indicates the
initial curve ($\phi= m_p$) that we use to find appropriate parameters
at $\bar{\mu}$, so that (by construction) we match observational
constraints at this value of $\phi$. The red star indicates
$\bar{\mu}$ where the initial curve ends, and all other curves (each
for a different value of $\phi$) begin. Blue dots indicate the
location of $\mu*$ on each curve.

The results are very similar to inflation. Smaller values of $\phi$
cause $\lambda_\phi({\mu*})$ to rapidly approach zero. Due to $M_\phi$
being proportional to an inverse power of $\phi$, decreasing $\phi$ also
increases $\mu* = M_\phi$. Thus the blue dots on each curve shift to the
right as $\phi$ is decreased, the opposite for inflation.

Making $\phi$ small enough eventually drives $\lambda_\phi(\mu*)$
past zero. Our concern for quintessence is different to that of
inflation; we are not concerned with $\lambda_\phi$ becoming
negative; rather we are concerned with it becoming $\emph{large}$
compared with the non-renormalizable term in the potential.  However,
once $\lambda_\phi(\mu*)$ becomes negative, smaller values of $\phi$
decrease it further, such that its absolute value becomes large
enough to be problematic.  Thus, to ensure the viability of the
quintessence model, it is sufficient to make the same demand as with
inflation: that $\lambda_\phi$ remains positive for a a considered
range in $\phi$.

\begin{figure}
\includegraphics[width=0.5\textwidth]{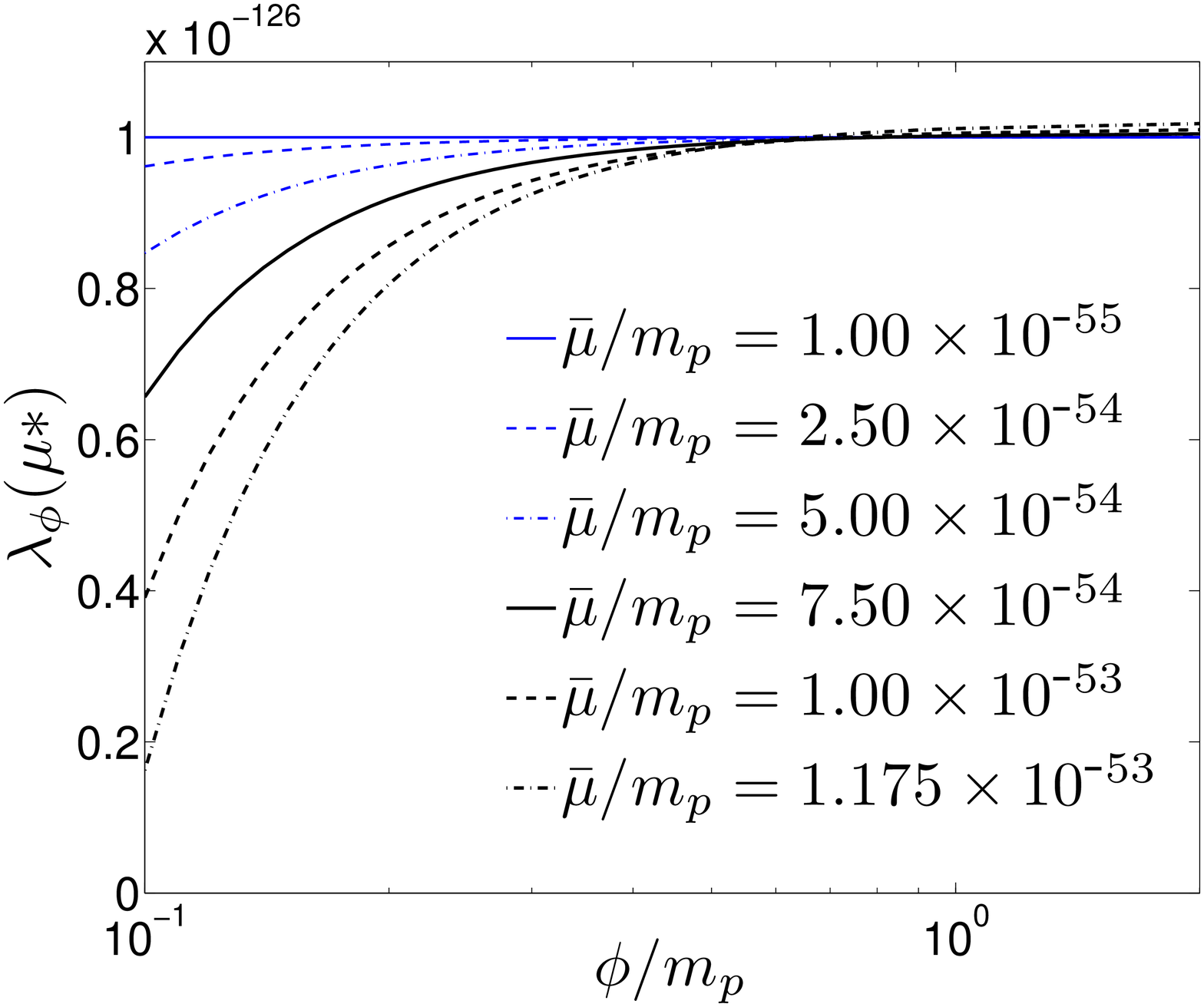}
\caption{\label{plot5}Curves of $\lambda_\phi(\mu*)$ against
  $\bar{\mu}$. The value of $g^2$ remains at $10^{-20}$. At large
  values of $\phi$ the curves become flat, while small values induce
  large changes in $\lambda_\phi(\mu*)$, eventually driving it
  negative (and eventually, large).  As with inflation, larger values
  of $g^2$ may be compensated for by using smaller values of
  $\bar{\mu}$, and vice-versa.}
\end{figure}

Just as before, we construct curves of $\lambda_\phi(\mu*)$ with
respect to $\phi$, shown on Fig. \ref{plot4}. As with inflation, very
small values of $\phi$ likely signal a breakdown in the description of
the field in terms of an effective potential. This is particularly
clear in the case of quintessence, where in the early universe and at
small field values $\phi \ll m_p$ the field may be moving extremely
quickly, to the extent that the field's behaviour is dominated by the
kinetic terms. Again at larger field values, there is little
difference to the value of the parameter.  Due to the
non-renormalizable term in the potential, smaller values of $\phi$
actually leads to a larger value of $\mu*$.  This makes little
difference to the behaviour of $\lambda_\phi(\mu*)$ with respect to
$\phi$ however, as the curves are so flat at low values of $\mu$.
Fig. \ref{plot5} shows this dependence of $\lambda_\phi(\mu*)$ with
$\phi$, for different values of $\bar{\mu}$. The behaviour is much the
same, and can be understood in precisely the same way.

The equations remain unchanged when moving from inflation to
quintessence.  The solution is therefore identical, and following the
same line of reasoning, we may therefore write down the constraint:
\be 
g \bar{\mu} \ll 10 M_\phi (\bar \mu) \simeq 10^{-42}\, \textrm{GeV} \,.  
\ee
Exactly how constrained the coupling $g^2$ is, depends upon the
renormalisation scale $\bar{\mu}$ at which the physics is set.

Quintessence is a low energy phenomena, operating at scales great many
orders of magnitude lower than inflation. The constraint on
$\lambda_\phi$ means that the combination $g \bar{\mu}$ must be much
smaller than in the case of inflation, to avoid generating corrections
to the effective potential that ruins the tree-level
behaviour. However once again there is a great deal of ambiguity as to
what an appropriate value of $\bar{\mu}$ should be. This is
information not included in quintessence models, just as it is not
included in models of inflation.

\section{Summary}
\label{sum}

The approach in this paper is consistent with the concepts of
low-energy effective field theories 
\cite{Polchinski:1983gv,bt,Burgess:2007pt}.  In
the effective field theory approach, at low energies, reflected
by a low renormalization scale, the effective theory is obtained
from the full theory by removing all heavy mass field internal
propagator lines and treating the heavy fields only
as external particles in Green's functions, with all other effects
emerging through renomalization of the parameters in the theory.
In this paper, these concepts have been applied to the effective
potential, in application to inflation and quintessence models.
This follows previous treatments of decoupling in effective potentials,
applied to electroweak physics \cite{bando2,nakano,casas}.
What is proposed here is that in interacting inflation or 
quintessence models, 
if there are disparate mass scales in the theory, then a mass
independent renormalization scheme, such as the 
${\rm MS}$ or ${\overline {\rm MS}}$
schemes is not adequate in capturing the physics of decoupling.
Rather the mass dependent renormalization scheme is more appropriate.
In this scheme, as shown in this paper, heavy fields coupled
to the inflaton or quintessence field have their quantum effects
power suppressed in the effective potential.  To implement the
MDR scheme, the choice of renormalization scale, 
and thus the division between low and
high energy scales in the effective potential, must be determined.

Although the effective potential must be independent of renormalization scale,
the issue is at what scale should the parameters in the theory
be initialized.  As noted in this paper, this is an underlying ambiguity
in inflation and quintessence models and can lead to differing
results.  We discussed two options which we called the low-energy 
and high-energy
approaches in Sect. \ref{results}. In the low-energy approach
the parameters are initialized at a renormalization scale at order the
lowest mass scale in the system.  One consequence of this approach
is, if the inflaton or quintessence field is the field with lowest mass,
and all other fields interacting with it are much heavier, then there
will be negligible quantum corrections, due to the decoupling
effects.  Alternative in the high-energy approach, the parameters are
initialized at some high energy scale, possibly if the theory were
embedded in some higher theory and some symmetry principle
determined the parameters at high scale. In this case, one can
use the renormalization group equations to evolve the scale
to where one wants to do physics.
Thus in the absence of the inflation
or quintessence model being embedded in some higher theory,
if the model has disparate mass scale, the low-energy approach is
technically a viable option, in which case even if the model is
moderately interacting with other fields, because of the
decoupling effect it can still produce
the very flat potentials needed.

In cosmology there has been a common practice in the treatment of the scalar
field effective potential, whereby the effect at one-loop
order of any quantum field coupled to this
scalar field leads to a Coleman-Weinberg correction term in
the quantum corrected effective potential \cite{Lyth:1998xn}.  
However this procedure is
not always applicable.  In particular, when the mass of a quantum
field is greatly in excess to all other masses in the system,
the decoupling theorem implies the quantum corrections
from this heavy field are suppressed in powers of the light-to-heavy
mass ratio.  

There are many common models in inflation
and quintessence where these suppression effects become valid.
For inflation, in all large field models, often terms chaotic inflation
models \cite{ci}, as well as some hybrid models \cite{Linde:1993cn}
inflation occurs when the inflaton background field 
value is
very large $\langle \phi \rangle \sim m_p$.  Thus any scalar or fermion
field coupled to the inflaton with at least moderate coupling will have
a very large mass, much bigger than the inflaton mass.  The quantum correction
from such fields will thus be greatly suppressed during the inflation 
period due to decoupling.
Also in small field models \cite{ni} after inflation, as the inflaton
background field value grows, it is possible fields coupled to the inflaton
acquire large mass and thus there quantum corrections might become
suppressed.  In quintessence models, the dark energy regime in most
models occurs when the scalar field background value
is very large $\phi \sim m_p$ 
\cite{quintessencep,couplequint2,Peebles:1998qn}.
Thus once again during the quintessence regime, quantum corrections
from other scalar and fermion fields coupled to the quintessence
field will be highly suppressed from decoupling effects.

This decoupling effect can have significant importance to the building
of inflation and quintessence models.  For inflation it implies the inflaton
field can be coupled more strongly to other fields.  Stronger
couplings can have beneficial effects in leading to more robust
reheating after inflation or in the case of 
warm inflation models \cite{BasteroGil:2009ec},
can lead to a greater parameter regime and many more possible
models.  Moreover inflation models in general can be less
dependent on supersymmetry.
In quintessence models, the potential during the
dark energy regime typically has to be so flat, that the quintessence
field often is just added as an additional field without any
other dynamical purpose aside from driving the late time
dark energy expansion of the universe.  However the decoupling effect
means that at early times the quintessence field might be
interacting with other fields and produce dynamical consequences
and at later time as the quintessence field background value becomes large,
these other fields become very massive, thus decoupling, which
then leads to a almost noninteracting field with the
required ultra-flat potential to drive the late time
dark energy phase.

\begin{acknowledgments}

M.B.G. is partially
supported by the M.E.C. under contract FIS2007-63364 and by the Junta
de Andaluc\'{\i}a group FQM 101.
A.B. and B.M.J. are supported by STFC.

\end{acknowledgments}

\section*{Appendix A: Mass Dependent Renormalization Scheme}
\label{appa}

In this Appendix we summarize the mass dependent renormalization scheme
\cite{mdrs}  used to obtain the RGEs for mass parameters and
couplings Eqs.~(\ref{betalam})-(\ref{betaL}). In order to set the
notation and the procedure, we start by considering
the simplest example, a single scalar field with potential: 
\be
V(\phi) = \frac{\lambda_\phi}{4!} \phi^4 + \frac{m_\phi^2}{2}\phi^2 \,.   
\ee
The renormalized Lagrangian including the counterterms is then:
\bea
{\cal L}_{\rm ren} ={\cal L} + \delta {\cal L}_{ct} &=& \frac{1}{2}
\partial_\mu \phi \partial^\mu \phi
-V(\phi) 
+ \frac{\delta_{Z_\phi} }{2} \partial_\mu \phi \partial^\mu \phi
- \frac{\delta m_\phi^2}{2}\phi^2 - \frac{\delta \lambda_\phi}{4!}
\phi^4 - \delta \Omega \nonumber \\ 
&=& \frac{Z_\phi}{2} \partial_\mu \phi \partial^\mu \phi
-\frac{Z_{m^2_\phi}^{-1}}{2}m^2_\phi \phi^2 - \frac{Z_\lambda }{4!} \lambda_{\phi} \phi^4 - Z_\Omega^{-1}\Omega \,, 
\label{lag}
\eea
where $\phi$, $m^2_{\phi}$ and $\lambda_{\phi}$ are (finite) 
renormalized parameters, and $\phi_0=Z_\phi^{1/2} \phi$,
$m^2_{\phi 0}= Z_\phi^{-1} Z_{m^2_\phi}^{-1} m^2_\phi$,
$\lambda_{\phi 0}= Z_\phi^{-2} Z_{\lambda_\phi} \lambda_\phi$,  the
(infinite) bare parameters, and 
\be
\delta_{Z_\phi}=Z_\phi-1 \,,
\ee
$Z_\phi$ being  the wave function renormalization constant. 

Loop corrections  are computed
with the Lagrangian given by the first line in Eq. (\ref{lag}),
including the corrections given by the counterterms. 
The first step is then to regularize
the divergent integrals, and for that we use dimensional
regularization: evaluating the integrals in $d=4 -\epsilon$ dimensions,
and then taking the limit $\epsilon$ going to zero. The divergent term
at one-loop is isolated as the single pole when $d=4$  ($2/\epsilon$ term).
The renormalized (finite) mass and coupling are defined by 
imposing suitable normalization conditions on the n-point 1PI Green
functions $\Gamma^{(n)}$ at some arbitrary scale $\mu$ \cite{ramond}. 
The relation between the bare and renormalized n-point 1PI functions
is given by:
\be
\Gamma^{(n)}(p^2) \mid_{p^2=\mu^2}=
Z_\phi^{n/2}\Gamma_0^{(n)}(p^2)\mid_{p^2=\mu^2} \,,
\ee
where as before the subscript ``0'' denotes the bare quantity. 
The normalization condition  fixes the counterterms that cancel out
the divergent terms; this is equivalent to 
define the renormalization constants, given the relation
between these and the counterterms introduced
in section \ref{sect2},  Eqs.(\ref{deltam})-(\ref{deltaL}). We now derive them
explicitly in the MDR scheme.  
  
\begin{figure}
\includegraphics[width=0.6\textwidth]{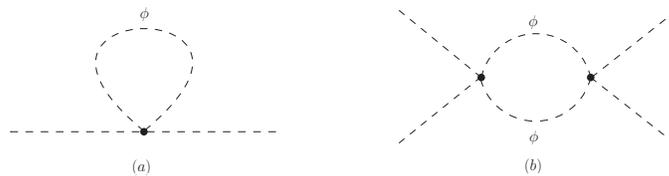}
\caption{\label{plot6} (a) One-loop scalar self-energy diagram. (b)
  One-loop correction to the proper scalar quartic vertex. }
\end{figure}

The 2-point 1PI renormalized function $\Gamma^{(2)}$,  including
the contributions from the counterterms and the radiative
correction $\Pi^{(\phi)}(p^2)$ (Fig. (\ref{plot6}.a)), is given by:   
\be
\Gamma^{(2)}(p^2) = p^2 - m_\phi^2 +
\delta_{Z_\phi}p^2 - \delta m_\phi^2 + \Pi^{(\phi)}(p^2) \,.   
\label{GammaPI2}
\ee
The counterterms, or equivalently the renormalization constants, are
fixed by demanding $\Gamma^{(2)}$ to be that of a free-field theory with
running mass parameter $m^2_{\phi}(\mu)$ ~\cite{decoupling},  at the
renormalization scale $\mu$: 
\be
\Gamma^{(2)}(p^2) \mid_{p^2= \mu^2} 
\equiv (p^2 - m^2_{\phi}(p^2))\mid_{p^2= \mu^2} \,, 
\label{Gamman2}
\ee
where $p^2$ is the incoming euclidean momentum\footnote{In this Appendix
  $p^2$ will refer to the Euclidean momentum, and we drop for
  simplicity the subindex 
  ``E'' hereon.}.  
The one-loop contribution is given in this
case by: 
\bea
\Pi^{(\phi)}(p^2) &=& \frac{\lambda_\phi}{2} L(m_\phi^2/\hat \mu^2) m_\phi^2 \,,\\
L(m_\phi^2/\hat \mu^2)&=& \frac{1}{16 \pi^2}\left(
\frac{2}{\bar\epsilon} + 1 - \ln \frac{m_\phi^2}{\hat \mu^2} \right) \,,
\eea
where the scale $\hat \mu$ is introduced in the regularization
procedure because of dimensional reasons, and $2/ \bar\epsilon=
2/\epsilon -\gamma_E + \ln 4 \pi$. When the scalar field does
not couple to fermions, the wave function renormalization constant at
1-loop does not receive any contribution and therefore:
\be
Z_\phi =1 \,.
\ee
On the other hand, the normalization condition fixes the mass
counterterm: 
\be
\delta m^2_{\phi} = \Pi^{(\phi)}(\mu^2) \,,
\ee
and from  Eq. (\ref{deltam}) we can read the mass renormalization
constant: 
\be
Z_{m_\phi^2}= 1 - \frac{\lambda_\phi}{2}L(m_\phi^2/\hat \mu^2) \,. 
\ee
The relation between the renormalized mass parameter and the bare
parameter is given by:  
\be
m_{\phi}^2( \mu) =Z_\phi Z_{m_\phi^2} m_{\phi 0}^2\,, 
\ee
and thus, taking the derivative with respect to the renormalization
scale, one obtains the RGE for the mass parameter: 
\be
\beta_{m_\phi^2}= \frac{d m^2_{\phi}( \mu)}{d \ln  \mu}=
m^2_{\phi} \frac{d \ln Z_{m_\phi^2}}{d \ln  \mu}=0 \,. 
\ee
Thus, given that at one-loop the radiative correction
$\Pi^{(\phi)}(p^2)$ is independent of the 
external momentum, this (quadratically) divergent contribution can be
reabsorbed into a redefinition of the mass parameter. 

For the renormalized vacuum energy, the situation is quite
similar. The zero-point 1PI function at 1-loop are given by vacuum
diagrams which do not depend on any scale except that of the mass of
the particle  in the loop. Therefore,  
\be
Z_\Omega \Omega = \Omega - \frac{m_\phi^4}{4}L(m_\phi^2/\hat \mu^2) \,,
\ee
and
\be
\beta_{\Omega}= 0 \,.
\ee
The vacuum contribution will be fixed by the
boundary conditions, say $\Omega(\mu)=0$. Either at higher orders, or when
the scalar couple to fermions, we will have $Z_\phi \neq 1$ and
$\beta_{m_\phi^2} \neq 0$, but still the pure vacuum contributions
give $\beta_{\Omega}= 0$.   

The RGE for the quartic coupling is derived in a similar manner,
obtaining the coupling renormalization constant by imposing the
normalization condition on the 4-point 1PI function. 
The renormalized 4-point 1PI function at 1-loop is given by: 
\be
\Gamma^{(4)}(p^2) = - \lambda_{\phi} - \delta \lambda_\phi +
\frac{3}{2} \lambda_\phi^2  \Gamma(p^2,m_\phi^2)
\ee
where $\Gamma(p^2,m^2)$ is the contribution from the 1-loop diagram
Fig. (\ref{plot6}.b):   
\be
\Gamma(p^2,m^2)= \frac{1}{16 \pi^2} \left( \frac{2}{\bar\epsilon} -
\int_0^1 dx \ln(\frac{m^2 + x(1-x) p^2}{\hat \mu^2}) \right) \,.
\label{Gamma}
\ee
We impose the normalization condition: 
\be
\Gamma^{(4)}(p^2)\mid _{p^2= \mu^2} \equiv -\lambda_{\phi}( \mu) \,,
\label{Gamma4}
\ee
which defines the coupling counterterm:
\be
\delta \lambda_\phi = \frac{3}{2} \lambda_\phi^2\Gamma(\mu^2,m_\phi^2) \,,
\ee
and using Eq. (\ref{deltalambda}), the coupling renormalization constant
$Z_{\lambda_\phi}^{-1}$: 
\be
Z_{\lambda_\phi}^{-1} = 1 - \frac{3}{2} \lambda_\phi \Gamma( \mu^2,m_\phi^2) \,. 
\label{Zlambda}
\ee
The relation between renormalized and bare coupling is given by: 
\be
\lambda_{\phi}( \mu) = Z_\phi^2  Z_{\lambda_\phi}^{-1} \lambda_{\phi 0} \,,
\ee
and as before, taking the derivative the RGE for the coupling reads:
\bea
\frac{d \lambda_\phi(\mu)}{d \ln \mu} &=& \frac{3
  \lambda_\phi^2}{(4 \pi)^2}F_2(a_\phi) \,,\label{betalambda}\\ 
F_2(a_\phi)&=& 1 - \frac{2 a_\phi}{\sqrt{1 + 4 a_\phi}} \ln \frac{\sqrt{1 + 4
    a_\phi}+1}{\sqrt{1 + 4 a_\phi}-1} \,, 
\eea
where $a_\phi=m_\phi^2/\mu^2$. Threshold effects are included in the
effective coupling through the 
momentum dependence $p^2= \mu^2$ of the radiative
corrections\footnote{In practice, the 
running effective coupling can be obtained by taking the derivative of
the one-loop 1PI Green functions  with respect to the momentum, and
then replacing $p^2= \mu^2$.}, such that  the
coefficients of the RGEs are modulated by a threshold function
$F_2(a)$. The  latter reduces to one in
the massless limit, $a=0$, but it  goes to zero when $a=m^2/\mu^2$
goes to infinity.  That is, in the massless limit one recover
the same RGEs for the effective couplings than those computed in a
mass independent scheme, like the $\overline {\rm MS}$ scheme. But in
the opposite limit, for a heavy state with $a \gg 1$, the contribution
is more and more suppressed as the ratio $a$ increases.

Having set the scheme in the simplest model, we can extend it now to
the case of study, adding the second scalar field 
$\chi$, with potential:
\be
V(\phi, \chi) =  
\Omega + \frac{1}{2} m^2_{\phi} \phi^2 +\frac{\lambda_\phi}{4!} \phi^4 
                + \frac{1}{2} m_{\chi}^2 \chi^2 
                + \frac{\lambda_{\chi}}{4!} \chi^4   
                + \frac{1}{2} g^2 \chi^2 \phi^2 \,. 
\ee
Adding the corresponding counterterms, the renormalized 
Lagrangian for the scalar sector is then: 
\bea
\!\!\!\!\!{\cal L}_{S} &=& \frac{1}{2}\partial_\mu \phi \partial^\mu \phi 
+\frac{1}{2} \partial_\mu \chi \partial^\mu \chi
-V(\phi, \chi) \nonumber \\
& &\!\!\!\!\!\!\!\!\!\!\!\!\!\!\!\!\!\!\!\!
 + \frac{\delta_{Z_\phi} }{2} \partial_\mu \phi \partial^\mu \phi
+ \frac{\delta_{Z_\chi} }{2} \partial_\mu \chi \partial^\mu \chi
- \frac{\delta m_\phi^2}{2}\phi^2  - \frac{\delta m_\chi^2}{2}\chi^2  
- \frac{\delta \lambda_\phi}{4!} \phi^4 - \frac{\delta
  \lambda_\chi}{4!} \chi^4 - \frac{\delta g^2}{2} \phi^2 \chi^2 
- \delta \Omega \,.
\eea
We also consider Yukawa interactions between $\chi$ and $N_F$ fermions
$\Psi_\alpha$, with the fermionic Lagrangian  given by:
\bea 
{\cal L}_{F} &=& \bar \Psi_\alpha ( i\gamma^\mu\partial_\mu - m_f)
\Psi_\alpha - h \chi \bar \Psi_\alpha \Psi + \delta_{Z_f} \bar
\Psi_\alpha \Psi - \delta m_f \bar \Psi_\alpha \Psi - \delta h \chi
\bar \Psi_\alpha \Psi \,, 
\eea
where we have taken for simplicity a common mass $m_f$ and Yukawa
coupling $h$ for the fermions. 

\begin{figure}
\includegraphics[width=0.6\textwidth]{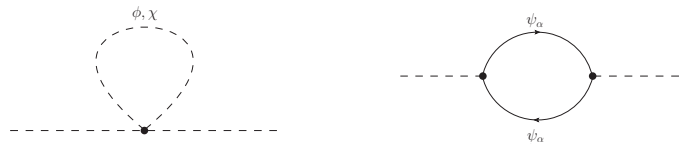}
\caption{\label{plot7} One-loop self-energy diagram contribution
  to the $\chi$ propagator. Dashed lines represents scalars, and
  fermions are given by solid lines.  }
\end{figure}

The interaction term given by
the coupling $g^2$ only adds a constant (quadratically) divergent term
to $\Pi^{\phi}$, through the same diagram than in Fig. (\ref{plot6}.a) but with
a $\chi$ internal line. Therefore, still we have $Z_\phi=1$ and  
\be
m_\phi^2Z_{m_\phi^2}= m_\phi^2 - \frac{\lambda_\phi}{2}
m_\phi^2L(m_\phi^2/\hat \mu^2) 
- g^2 m_\chi^2L(m_\chi^2/\hat \mu^2) \label{Zmphi}
\,,
\ee
and thus $\beta_{m^2_\phi}=0$. On the other hand,
$Z_\chi$ and $Z_{m_\chi^2}$ receive a $p$-dependent contribution from the loop of
fermions. The diagrams contributing to the $\chi$ field
2-point function are given in Fig. (\ref{plot7}), with 
\bea
\Pi^{(\chi)}(p^2) &=& 2 h^2 N_F \Gamma(p^2, m^2_f) p^2 + 
8 h^2 N_F \Gamma (p^2, m^2_f) m^2_f
+\frac{\lambda_\chi}{2} L(m_\chi^2/\hat \mu^2) m_\chi^2  
+g^2 L(m_\phi^2/\hat \mu^2) m_\phi^2 
 \,. 
\eea
For the couplings, the renormalization constants $Z_\chi$ and
$Z_{m_\chi^2}$ are given then by: 
\bea
Z_\chi    &=& 1 - 2 h^2 N_F \Gamma(\mu^2, m^2_f) \,,  \label{Zchi}\\
m_\chi^2Z_{m_\chi^2}&=& m_\chi^2 - 8 h^2 N_F m_f^2\Gamma(\mu^2, m^2_f) 
-\frac{\lambda_\chi}{2} m_\chi^2 L(m_\chi^2/\hat \mu^2) 
-g^2m_\phi^2 L(m_\phi^2/\hat \mu^2) 
\,.  
\label{Zmchi}
\eea

\begin{figure}
\includegraphics[width=0.6\textwidth]{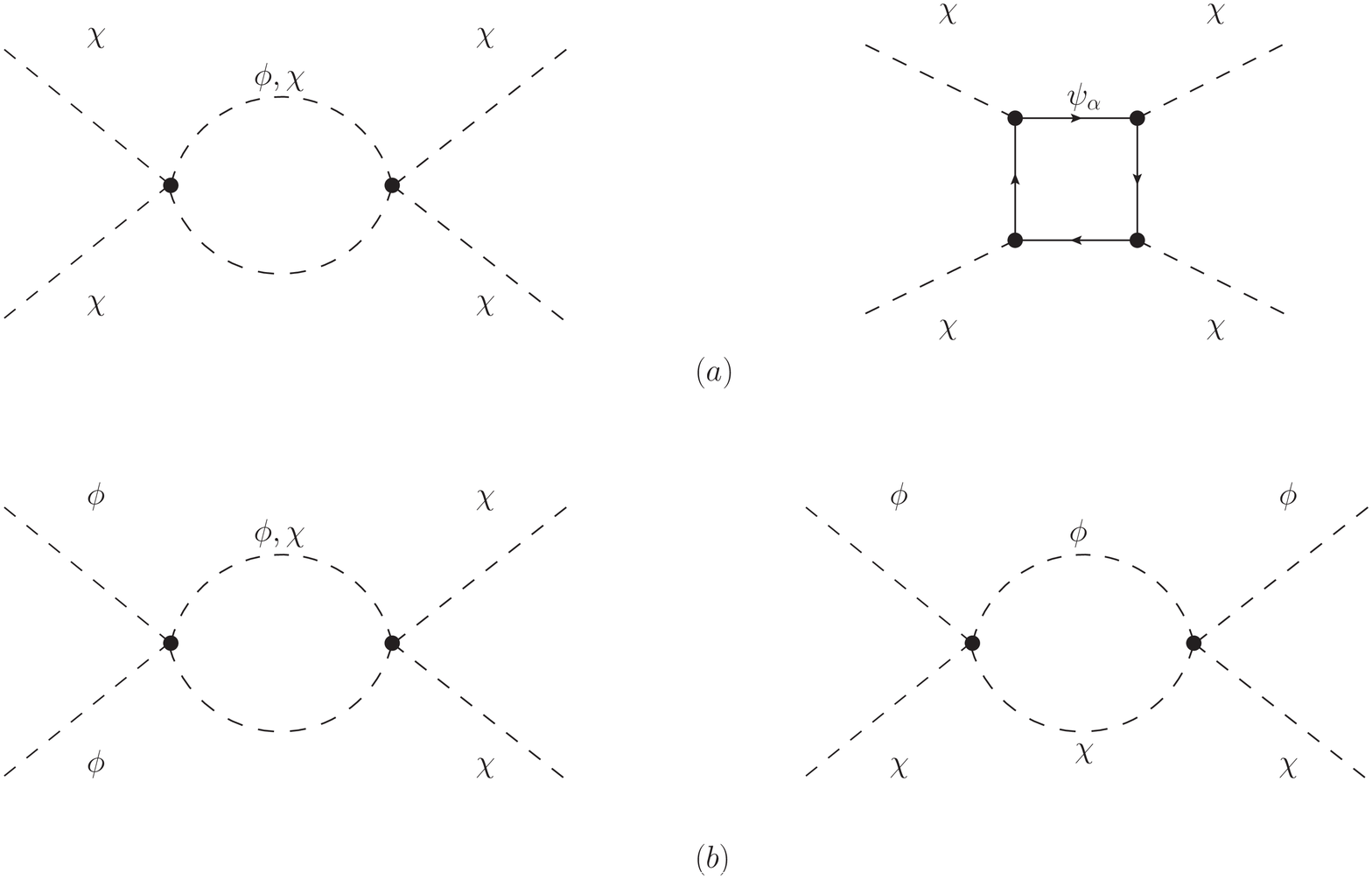}
\caption{\label{plot8} (a) One-loop proper vertex correction to
  $\lambda_\chi$. (b) One-loop proper vertex correction to $g^2$. }
\end{figure}

The renormalization constant $Z_{\lambda_\phi}^{-1}$ picks up a new term similar
to that in Eq. (\ref{Zlambda}) due to the loop with 2 massive $\chi$
states, i.e., a diagram similar to that in Fig. (\ref{plot6}.b) but with  $\chi$
running in the loop. The 1-loop diagrams contributing to the $\lambda_\chi$
and $g^2$ coupling renormalization constants are given in
Figs. (\ref{plot8}.a) and (\ref{plot8}.b) respectively.
The set of renormalization constant are then: 
\bea
\lambda_\phi Z_{\lambda_\phi}^{-1}&=& \lambda_\phi - \frac{3}{2} \lambda_\phi^2
\Gamma(\mu^2, m_\phi^2) - 6 g^4 \Gamma(\mu^2, m_\chi^2)
\,, \label{Zlambdaphi}\\
\lambda_\chi Z_{\lambda_\chi}^{-1}&=& \lambda_\chi - \frac{3}{2} \lambda_\chi^2
\Gamma(\mu^2, m_\chi^2) - 6 g^4 \Gamma(\mu^2, m_\phi^2)
- 24 h^4 N_F \Gamma( \mu^2, m_f^2)
\,, \label{Zlambdachi}\\
g^2 Z_{g^2}^{-1}&=& g^2\left(1 - \frac{\lambda_\chi}{2} \Gamma( \mu^2,
m_\chi^2)
-\frac{\lambda_\phi}{2} \Gamma(\mu^2, m_\phi^2)
 - 4 g^2 \Gamma_2(\mu^2, m_\chi^2, m_\phi^2)\right)\, , \label{Zg2}
\eea
where:
\be
\Gamma_2(p^2,m_1^2,m_2^2)= \frac{1}{16 \pi^2} \left( \frac{2}{
  \bar \epsilon} -\int_0^1 dx \ln(\frac{m_1^2 x + m_2^2(1-x) + x(1-x)
  p^2}{\hat \mu^2}) \right) \,. 
\label{Gamma2}
\ee

\begin{figure}
\includegraphics[width=0.5\textwidth]{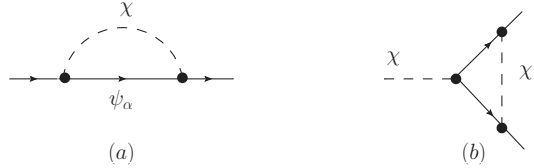}
\caption{\label{plot9} (a) One-loop fermion self-energy. 
(b) One-loop Yukawa vertex correction.}
\end{figure}

To obtain the RGE for the Yukawa coupling we also need  to renormalize
the inverse fermion propagator $S^{-1}$ and the Yukawa vertex
$\Gamma^{(3)}$. The corresponding 1-loop diagrams are given in
Figs. (\ref{plot9}.a) and (\ref{plot9}.b). The normalization condition
for $S^{-1}$  and $\Gamma^{(3)}$  are similar to Eqs. (\ref{Gamman2}) and
Eqs. (\ref{Gamma4}): 
\bea
S^{-1}(p^2)_{p^2= \mu^2}& \equiv& (p \!\!\! \slash -
m_f(p^2))\mid_{p^2= \mu^2} \,, \\ 
\Gamma^{(3)}(p^2)_{p^2= \mu^2}& \equiv& h_R( \mu) \,, 
\eea
and from the above equations we can obtain $Z_f$, $Z_h$, and $Z_{m_f}$
for massive fermions: 
\bea
Z_f &=& 1 - \frac{h^2}{2} \Gamma_f(\mu^2,m_\chi^2,m_f^2) \,, \\
Z_h^{-1} &=& 1  - h^2 \Gamma_2(\mu^2,m_\chi^2,m_f^2) \,, \\
Z_{m_f} &=&  1  - h^2 \Gamma_f(\mu^2,m_\chi^2,m_f^2) \,, 
\eea 
where:
\bea 
\Gamma_f(p^2,m_1^2,m_2^2)&=& \frac{1}{16 \pi^2} \left( \frac{2}{\bar\epsilon} -
\int_0^1 dx x \ln(\frac{m_1^2x + m_2^2 (1-x) + x(1-x) p^2}{\hat \mu^2})
\right) \,. 
\label{Gammaf}
\eea

Finally, from the above renormalization constant, and the relations:
\bea
m^2_{\phi}(\mu) &=& Z_\phi Z_{m^2_\phi} m^2_{\phi 0} \,,\\
m^2_{\chi}(\mu) &=& Z_\chi Z_{m^2_\chi} m^2_{\chi 0} \,,\\
\lambda_{\phi}( \mu) &=& Z_\phi^2  Z_{\lambda_\phi}^{-1}
\lambda_{\phi 0} \,,\label{lamphir}\\
\lambda_{\chi}( \mu) &=& Z_\chi^2  Z_{\lambda_\chi}^{-1}
\lambda_{\chi 0} \,,\\
g^2(\mu) &=& Z_\phi Z_\chi   Z_{g^2}^{-1} g^2_{0} \,,\\
h^2(\mu) &=& Z_\chi^{1/2} Z_f   Z_{h^2}^{-1} h^2_{0} \,,
\eea
the  RGEs  (\ref{betalam})-(\ref{betam}) including the threshold functions
are easily derived (setting $m_f=0$). The latter are given by:    
\bea
F_1(a)&=& -(4\pi)^2 \frac{d \Gamma_2(p^2,m^2,0)}{d \ln p^2}=
1- a \ln \frac{1+ a}{a} \,, \label{F1}\\
F_2(a)&=& -(4\pi)^2 \frac{d \Gamma(p^2,m^2)}{d \ln p^2}=
1 - \frac{2 a}{\sqrt{1 + 4 a}} \ln \frac{\sqrt{1 + 4
    a}+1}{\sqrt{1 + 4 a}-1} \,, \label{F2}\\
F_3(a) &=& -(4\pi)^2 \left(\frac{d \Gamma_f(p^2,m^2,0)}{d \ln p^2}
+ 2\frac{d \Gamma_2(p^2,m^2,0)}{d \ln p^2} \right)=
1 + 2 a ( 1 - (1+ a) \ln \frac{1+ a}{a}) \label{G1}\,,
\eea
where in computing $F_1(a)$ from the loop with one light $m_\phi$ and a
heavy $m_\chi \gg m_\phi$ we have set $m_\phi=0$. 
$F_2(a)$ is the threshold function for a scalar loop with 2 equal
massive states, $F_1(a)$ that of a scalar loop with one massless and
one massive scalar state, and $F_3(a)$ that with massless
fermions and one massive scalar.

\section*{Appendix B: power suppression of two-loops coefficients}

At two-loop order, the RG-improved effective potential is given by the
one-loop effective potential with running parameters evaluated using
the two-loop RG equations. Following the MDR prescription, again the
optimal choice to fix the renormalization scale is below all massive
thresholds; thus the two-loop effective potential reduces to the
tree-level potential plus the two-loop RGE functions. Decoupling will be
included in the latter through threshold functions, similarly to the
one-loop RGEs. 

\begin{figure}
\includegraphics[width=0.77\textwidth]{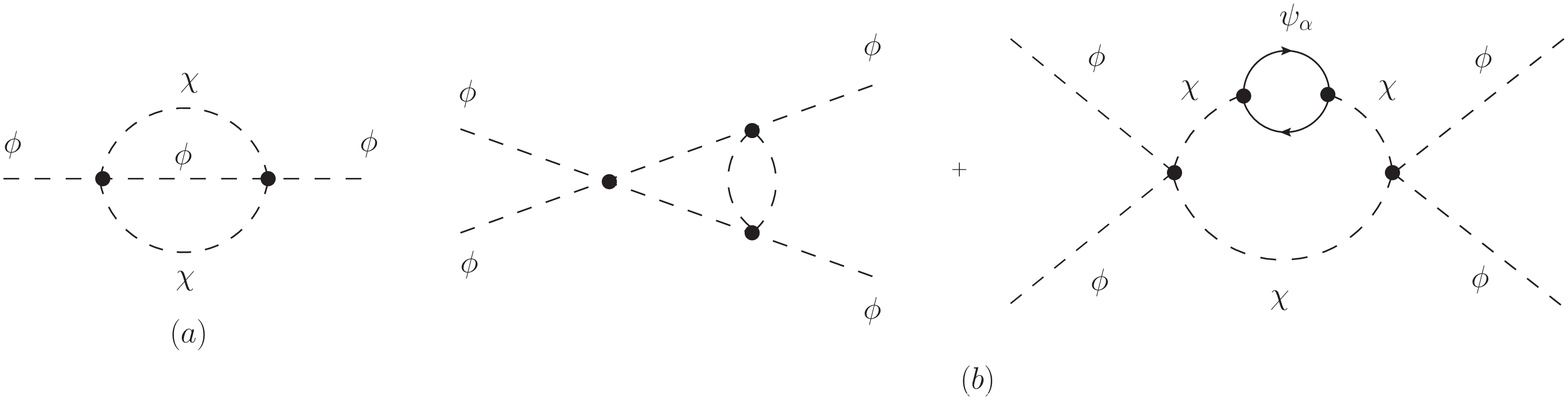}
\caption{\label{plot10} (a) Two-loops scalar wave function
  renormalization diagram.(b) Two-loops correction to the proper
  scalar quartic vertex. }  
\end{figure}

At two-loop order, both the wave function renormalization constant
$Z_\phi$ and the mass parameter one $Z_{m_\phi^2}$ get a $\mu$
dependent contribution from the sunset diagram in Fig. (\ref{plot10}.a),
whereas the quartic coupling $\lambda_\phi$ correction comes from
Fig. (\ref{plot10}.b). The two-loop beta functions in a mass
independent scheme can be found for example in Refs. \cite{2loops,2loopV},
where one only would need to extract the divergent contributions from those
diagrams. In the MDR scheme we need to carry out the full calculation of the
diagram keeping the finite contributions. We will not attempt such a
full two-loop calculation here, and we only want to argue that such
diagrams gives a power suppression of the corresponding RGE
coefficients that go at least like $O(\mu^2/M_\alpha^2)$ when $M_\alpha^2 \gg
\mu^2$, $M_\alpha^2$ being the heavy mass running in the loop. For the
first diagram  in the vertex correction in Fig.(\ref{plot10}.b), this
can be viewed as the 1-loop vertex correction but with the LHS
interaction replaced by an effective (momentum dependent) vertex again
like that of Fig. (\ref{plot6}.b). According to the one-loop
calculation, this  
will give the corresponding suppression
by the mass running in the loop also in the two-loop coefficients. 
The contributions of the order of
$O(\lambda_\phi g^4)$, and $O(g^6)$, will be therefore suppressed 
by the heavy mass $M_\chi^2$. That of the order of $O(\lambda_\phi^3)$
is suppressed by factors $O(\mu^2/M_\phi^2)$ when $\mu \ll M_\phi$. 
In the cosmology models studied in the text of this paper,
generally $M_\phi \sim \mu$ and so these diagrams are not suppressed
due to decoupling effects, but rather because $\lambda_\phi$ is
always tiny. The second diagram in Fig. (\ref{plot10}.b) gives a term
$(O(h^2 g^4)$ coming from the insertion of the fermion loop in one of
the internal $\chi$ lines. Again, this contribution will be suppressed
by a factor $O(\mu^2/M_\chi^2$, similarly to the 1-loop vertex
correction without the fermion insertion.

We will therefore concentrate on the sunset diagram in Fig. (\ref{plot10}.a),
and in particular on the contribution to the wave function
renormalization constant. In general for the sunset diagram with three
different internal scalar 
masses we have:
\bea
\Pi^{\phi}_{S}(p^2, m_1^2, m_2^2,m_3^2)\!  &=&\!  \frac{\lambda_i^2}{6} \left(
\sum_{j=1,2,3} m_j^2 A_j(p^2,m_i^2) + p^2 B(p^2,m_i^2)
\right) \,, \\
(4 \pi)^4 A_j(p^2,m_i^2)\!  &=& \! -2 ( \frac{1}{\epsilon^2} 
+ \frac{1}{\epsilon} (
\frac{3}{2}- \ln \frac{m_j^2}{4 \pi \hat \mu^2} -\gamma_E +
I_m(p^2,m_i^2) )  \,, \\
(4 \pi)^4 B(p^2,m_i^2)\! &=& \! \frac{1}{2} ( \frac{1}{\epsilon} -\gamma_E +
\frac{1}{2} - 2 \int_0^1 \!\!\! dx \int_0^1 \!\!\!dy (1-y) \ln (M^2(x,y) + p^2 y
(1-y)) ) \,, 
\eea
where  $M^2(x,y)= (m_1^2 x + m_2^2(1-x))y/(x(1-x)) + m_3^2 (1-y)$, and
$\lambda_i$ is a general quartic coupling with the appropriate symmetry
factors. In particular, when we have
three $\phi$ running in the loop then $\lambda^2_i= \lambda^2_\phi$, and with
two $\chi$ and one $\phi$ then $\lambda^2_i= 12 g^4$. The function
$I_m(p^2,m_i^2)$ is the finite contribution to the mass
renormalization. By adding the sunset 
contribution  to the 2-point function Eq. (\ref{GammaPI2}),
the normalization condition Eq. (\ref{Gamman2}) fixes the
non-vanishing two-loop wave
function counterterm $\delta_Z^{(2)}$, which defines the wave function
renormalization constant $Z_\phi^{(2)} = 1 +\delta_Z^{(2)}$, and then
the anomalous dimension of the field:
\be
\gamma^{(2)}_\phi = \frac{d \ln Z_\phi^{(2)}}{ d \ln \mu^2}= 
\frac{1}{(4 \pi)^4} ( \frac{\lambda_\phi^2}{12}
F_{22}(a_\phi,a_\phi,a_\phi) + g^4 F_{22} (a_\chi,a_\chi,a_\phi) ) \,,
\ee
with the threshold function given by the parametric integral:
\be
F_{22}(a_1,a_2,a_3)= 2 \int_0^1 dx \int_0^1 dy \frac{y (1-y)^2}{ a(x,y) + y (1-y)} \,,
\ee
where $a(x,y)=M^2(x,y)/\mu^2$. From the above expression, one can check
that indeed when we have only massless states running in the loop then
$F_{22}(0)=1$, but when any of the masses $a_i$ goes to infinity, then
$F_{22}(a_i)$  goes to zero as $O(1/a_i)$. The integral can be evaluated
numerically for arbitrary mass parameters, and it behaves similarly to
the 1-loop threshold functions in Fig. (\ref{plot1}). 

From the renormalization condition Eq. (\ref{lamphir}), the two-loop
beta function for $\lambda_\phi$ gets contributions from the
anomalous dimension $\gamma_\phi^{(2)}$ and the proper vertex
$\gamma_V^{(2)} = -d \ln Z_{\lambda_\phi}^{(2)}/ d \ln \mu$,
\bea
\beta_{\lambda_\phi}^{(2)} &=& 4 \lambda_\phi \gamma_\phi^{(2)} +
\gamma_V^{(2)} \nonumber \\   
&=& \frac{1}{(4 \pi)^4} \left( \lambda_\phi^3 ( \frac{1}{3}
F_{22}(a_\phi,a_\phi,a_\phi) - 6 G(a_\phi,a_\phi))\right. \nonumber \\
& & \left. \hspace{-0.47cm}  +  \lambda_\phi g^4 (4 F_{22} (a_\chi,a_\chi,a_\phi)
- 24 G(a_\phi,a_\chi) - 96 g^6 G(a_\chi, a_\chi) - 432 N_F h^2 g^4
G_F(a_f, a_\chi)\right) .  
\eea
We have not computed explicitly the 2-loop vertex threshold functions
$G(a_i,a_j)$, $G_F(a_i,a_h)$, but as argued above, we can expect them
to have the correct limits when $a_i \gg 1$, and therefore decoupling
in the sense of power suppression is also maintained in the MDR scheme
at the 2-loop level. 

\section*{Appendix C: 1-loop effective potential within the mass dependent renormalization (MDR) scheme}

The 1-loop radiative correction to the effective potential, computed
using dimensional regularization,  
is given by:
\be
\Delta V^{(1)}_{reg}= -\frac{1}{64 \pi^2} 
\left (M_\phi^4 (L(M_\phi^2/\hat \mu^2) + \frac{1}{2}) + 
M_\chi^4 (L(M_\chi^2/\hat \mu^2) + \frac{1}{2}) \right) \,.
\ee
The renormalized 1-loop effective potential is obtained by adding to
$\Delta V^{(1)}_{reg}$ the appropriate counterterms:
\be
\Delta V^{(1)}=\Delta V^{(1)}_{reg} + (Z_{\Omega}^{-1}-1) 
\Omega+ \frac{1}{2} m_\phi^2 (
Z_{m_\phi^2}^{-1}-1) \phi^2  
+ \frac{\lambda_\phi}{4!} (Z_{\lambda_\phi}-1) \phi^4 \,,
\ee
and plugging $Z_\phi=1$ and the renormalization constants given in
Eqs. (\ref{Zmphi}), 
(\ref{Zlambdaphi}), computed in the
MDR scheme, one has\footnote{The cosmological constant is renormalized
  by demanding $V(\phi=0)=0$.}:
\bea
\Delta V^{(1)}&=& \frac{1}{64 \pi^2} \left( 
m_\phi^4 \ln \frac{M_\phi^2}{m_\phi^2}
+\lambda_\phi m_\phi^2 \phi^2 \ln \frac{M_\phi^2}{m_\phi^2} 
+m_\chi^4 \ln \frac{M_\chi^2}{m_\chi^2}
+2 g^2 m_\chi^2 \phi^2  \ln \frac{M_\chi^2}{m_\chi^2}  \right. \nonumber \\
& & \left. + \frac{\lambda_\phi^2 \phi^4}{4}( \ln \frac{M_\phi^2}{
  \mu^2} -I(m_\phi^2/ \mu^2)  )       
+ g^4 \phi^4( \ln \frac{M_\chi^2}{
  \mu^2} -I(m_\chi^2/ \mu^2)) \right) \,,
\label{Veffmds}
\eea       
where 
\be
I(a)= \int_0^1 dx \ln(a + x(1-x)) = \ln a -2 -\sqrt{1 + 4 a } \ln
\frac{\sqrt{1+ 4a} -1}{\sqrt{1+ 4a}+1} \,.
\ee
Because the one-loop renormalization constant have been computed in
the previous section in 
the symmetric phase of the theory (i.e., taking the vev of the field
to vanish), the threshold functions $I(a)$
depend on the mass parameters $m_i^2$, instead of the physical
effective  masses $M_i^2$ relevant for the effective potential. Then, 
this expression would only 
lead to decoupling of heavy states when $m_\phi \,, m_\chi \gg
 \phi$, but not when $\phi \gg  m_\phi,\, m_\chi$. For example by
 taking $\mu \ll m_\phi,\, m_\chi$, and expanding the threshold
 function $I(a)$ when $a \gg 1$, we have:
\bea
\Delta V^{(1)}&=& \frac{1}{64 \pi^2} \left( 
M_\phi^4 \ln \frac{M_\phi^2}{m_\phi^2}
+M_\chi^4 \ln \frac{M_\chi^2}{m_\chi^2}
  \right. \nonumber \\
& & \left. + \frac{\lambda_\phi^2 \phi^4}{4}( -\frac{\mu^2}{6 m_\phi^2}
  + \cdots  )        
+ g^4 \phi^4(-\frac{\mu^2}{6 m_\chi^2}  + \cdots   ) \right) \,,
\eea       
and unless $m_\phi \sim m_\chi$, we are still left with potentially
large logs, $\ln \phi^2/m_\phi^2$, $\ln \phi^2/m_\chi^2$, and the
original problem addressed in Refs. \cite{bando2,kastening}.  
On the other hand, in the particular limit that the mass 
parameters vanish,  $m_\phi=m_\chi=0$,  one just recover
the expression for the effective potential computed in a mass
independent scheme: 
\be
\Delta V^{(1)}= \frac{1}{64 \pi^2} \left( 
\frac{\lambda_\phi^2 \phi^4}{4}( \ln \frac{\lambda_\phi \phi^2/2}{
  \mu^2} - 2 )       
+ g^4 \phi^4( \ln \frac{g^2\phi^2}{
  \mu^2} - 2 ) \right) \,.
\ee       
In this case, given that both mass scales are set by the vev of the field
$\phi$, large logs could be controlled and resummed by taking for example $
\mu \simeq \phi$. 

This apparent failure of the MDR scheme can be related to the fact
that the effective potential
is computed in the non-symmetric phase of the theory,  after shifting the
field $\phi$ by its vev. One should therefore
also impose the renormalization conditions and get the counterterms in
this phase.  After shifting the field, the propagators running
in the loops depend now on the effective mass and, by repeating the
calculation done in the previous section, one can derive similarly the
renormalization constants now with the threshold functions
depending on $M_\alpha$. Therefore, this 
accounts to replace  $m_i^2$ by  $M_i^2$ in both the logs and threshold
functions in Eq. (\ref{Veffmds}), and the effective potential is then
given by: 
\bea
\Delta V^{(1)} &=& \frac{1}{64 \pi^2} \left( 
\frac{\lambda_\phi^2 \phi^4}{4}( \ln \frac{M_\phi^2}{
  \mu^2} -I(M_\phi^2/ \mu^2))       
+ g^4 \phi^4( \ln \frac{M_\chi^2}{
  \mu^2} -I(M_\chi^2/ \mu^2)) \right) \,.  
\eea       
This is similar to the effective potential obtained in
Ref. \cite{casas}. In that work, decoupling is introduced in the
effective potential through step-functions at each physical threshold
$M_i=\mu$, Eq. (\ref{Vcasas}). In our case it is implemented through
the threshold function $I(a_i)$ obtained when computing the 1-loop
radiative corrections. 

Finally, the RG-improved effective potential is given by absorbing the
log dependence on the running parameters, such that the 1-loop
potential is given by the tree-level potential evaluated at the
boundary $t_*=\ln \mu_*/\mu$: 
\bea
V^{eff}&=& \frac{1}{2} m^2_{\phi}(t_*) \phi^2(t_*)
+\frac{\lambda_\phi(t_*)}{4!} \phi^4(t_*) \label{Veffappend} 
\,.  
\eea
This can be shown by explicit integration of the RGEs, plugging the
result in the tree-level potential. At 1-loop order, the mass
parameter and the field do not run, $m_\phi(t_*)= m_\phi$, and
$\phi(t_*)=\phi$, and integrating the RGE
for $\lambda_\phi$ in the MDR scheme, Eq. (\ref{betalam})  we have: 
\bea
\lambda_\phi(\mu_*)&\simeq& \lambda_\phi(\mu)  + \frac{3 \lambda^2_\phi(\mu)}{32
  \pi^2} \left( I(M_\phi^2/\mu_*^2) - I(M_\phi^2/\mu^2) - \ln
\frac{\mu_*^2}{\mu^2} \right) \nonumber \\
&& + \frac{12 g^4(\mu)}{32
  \pi^2} \left( I(M_\chi^2/\mu_*^2) - I(M_\chi^2/\mu^2) - \ln
\frac{\mu_*^2}{\mu^2} 
 \right) \,.
\eea
Taking $\mu_* \ll M_\phi,\, M_\chi$, this gives:
\be
\lambda_\phi(\mu_*) \simeq  \lambda_\phi(\mu) + 
\frac{3 \lambda_\phi^2(\mu)}{32\pi^2} \left( 
\ln\frac{M_\phi^2}{\mu^2} - I(M_\phi^2/\mu^2)\right)
+\frac{12 g^4(\mu)}{32\pi^2} \left( 
\ln\frac{M_\chi^2}{\mu^2} - I(M_\chi^2/\mu^2)\right) \,,
\ee
and therefore: 
\bea
V^{eff} &=& \frac{1}{2} m^2_{\phi}(t_*) \phi^2(t_*)
+\frac{\lambda_\phi(t_*)}{4!} \phi^4(t_*)   \nonumber \\
&\simeq& \frac{1}{2} m^2_{\phi} \phi^2+\frac{\lambda_\phi}{4!} \phi^4
\nonumber \\ 
&&+\frac{1}{64 \pi^2} \left( 
\frac{\lambda_\phi^2 \phi^4}{4}( \ln \frac{M_\phi^2}{
  \mu^2} -I(M_\phi^2/ \mu^2))       
+ g^4 \phi^4( \ln \frac{M_\chi^2}{
  \mu^2} -I(M_\chi^2/ \mu^2))\right) \,,
\eea
where  $\lambda_\phi \equiv \lambda_\phi(\mu)$ and
$g^2 \equiv g^2(\mu)$.

\end{document}